\documentclass[conference,compsoc]{IEEEtran}
\IEEEoverridecommandlockouts % needed for \thanks (corresponding-author note)
\ifCLASSOPTIONcompsoc
  \usepackage[nocompress]{cite}
\else
  \usepackage{cite}
\fi
\ifCLASSINFOpdf
  \usepackage[pdftex]{graphicx}
  \graphicspath{{figures/}}
\else
\fi
\ifCLASSOPTIONcompsoc
 \usepackage[caption=false,font=footnotesize,labelfont=sf,textfont=sf]{subfig}
\else
 \usepackage[caption=false,font=footnotesize]{subfig}
\fi
\usepackage[most]{tcolorbox}
\usepackage{tabularx}
\usepackage{booktabs}
\usepackage{enumitem}
\usepackage{amsmath}
\usepackage{soul}
\usepackage{amssymb}
\usepackage{array}
\usepackage{listings}
\usepackage{xcolor}
\usepackage{algorithm2e}
\usepackage{threeparttable}
\usepackage{pifont}

\lstdefinestyle{skillcode}{
  basicstyle=\ttfamily\small,
  breaklines=true,
  columns=fullflexible,
  keepspaces=true,
  showstringspaces=false,
  frame=single,
  rulecolor=\color{black!35},
  backgroundcolor=\color{gray!5},
  xleftmargin=1pt,
  xrightmargin=1pt,
  framexleftmargin=2pt,
  framexrightmargin=2pt,
  aboveskip=0pt,
  belowskip=0pt
}

\usepackage{tikz}
\newcommand{\circled}[1]{\tikz[baseline=(char.base)]{
  \node[shape=circle,fill=black,text=white,inner sep=1pt,
        font=\scriptsize\bfseries] (char) {#1};}}

\newcommand{\descr}[1]{\vspace{0.07in}\noindent\textbf{#1}}
\newcommand{\descrit}[1]{\vspace{0.07in}\noindent\emph{#1}}

\newcommand{\name}{\textsc{SelfOp}}
\newcommand{\cybergym}{\textsc{CyberGym}}

\newcommand{\rev}[1]{{#1}}
\newcommand{\revon}{}

\usepackage{xcolor}
\definecolor{d4blue}{HTML}{4C72B0}
\definecolor{d5loose}{HTML}{55A868}
\definecolor{d5default}{HTML}{DD8452}
\definecolor{d5strict}{HTML}{C44E52}
\definecolor{d7annot}{HTML}{8172B3}

\newcommand{\plotlegend}[2]{%
  \raisebox{-0.15ex}{\colorbox{#1}{\rule{0.55em}{0.45em}}}~#2%
}

\begin{document}
\title{\name: An Optimization Algorithm for Self-Improving Security Agents}

\author{
\IEEEauthorblockN{\rev{Saad Ullah$^{\star}$\thanks{\rev{$^{\star}$Corresponding author.}}}}
\IEEEauthorblockA{\rev{ECE Department}\\
\rev{Boston University}\\
\rev{saadu@bu.edu}}
\and
\IEEEauthorblockN{\rev{Yigitcan Kaya}}
\IEEEauthorblockA{\rev{CS Department}\\
\rev{UC Santa Barbara}\\
\rev{yigitcan@ucsb.edu}}
\and
\IEEEauthorblockN{\rev{Christopher Kruegel}}
\IEEEauthorblockA{\rev{CS Department}\\
\rev{UC Santa Barbara}\\
\rev{chris@cs.ucsb.edu}}
\and
\IEEEauthorblockN{\rev{Giovanni Vigna}}
\IEEEauthorblockA{\rev{CS Department}\\
\rev{UC Santa Barbara}\\
\rev{vigna@cs.ucsb.edu}}
\and
\IEEEauthorblockN{\rev{Gianluca Stringhini}}
\IEEEauthorblockA{\rev{ECE Department}\\
\rev{Boston University}\\
\rev{gian@bu.edu}}
}


% compsoc conferences compared to traditional conferences), use this
% \author{\IEEEauthorblockN{Michael Shell\IEEEauthorrefmark{1},
% Homer Simpson\IEEEauthorrefmark{2},
% James Kirk\IEEEauthorrefmark{3}, 
% Montgomery Scott\IEEEauthorrefmark{3} and
% Eldon Tyrell\IEEEauthorrefmark{4}}
% \IEEEauthorblockA{\IEEEauthorrefmark{1}School of Electrical and Computer Engineering\\
% Georgia Institute of Technology,
% Atlanta, Georgia 30332--0250\\ Email: see http://www.michaelshell.org/contact.html}
% \IEEEauthorblockA{\IEEEauthorrefmark{2}Twentieth Century Fox, Springfield, USA\\
% Email: homer@thesimpsons.com}
% \IEEEauthorblockA{\IEEEauthorrefmark{3}Starfleet Academy, San Francisco, California 96678-2391\\
% Telephone: (800) 555--1212, Fax: (888) 555--1212}
% \IEEEauthorblockA{\IEEEauthorrefmark{4}Tyrell Inc., 123 Replicant Street, Los Angeles, California 90210--4321}}


\maketitle

\begin{abstract}
LLM agents are increasingly used for security tasks such as vulnerability discovery, exploit reproduction, and patch generation.
Recent work has improved these agents by strengthening the underlying model, the execution harness, or the task context they consume.
However, model-level improvement typically requires either large supervised datasets, such as expert demonstrations or teacher-agent traces, or diverse environments with computable reinforcement rewards.
For security tasks, these resources are scarce: successful traces are costly to obtain, failures are difficult to diagnose, and rewards are often sparse, delayed, or non-computable.
Consequently, many efforts instead optimize the agent's harness or context.
Manual optimization of harness or context requires task-specific expertise and scales poorly, while automated methods often depend on scarce ground truth, stronger optimizer models, or unguided propose-and-evaluate loops that reduce optimization to unreliable and costly trial and error.

We introduce \name, an optimization algorithm for automatically improving a frozen security agent's task context, including instructions, skills, and reference documents, without modifying the agent's execution harness or underlying model weights.
\name{} casts context optimization as chain-rule-inspired textual gradient descent.
Starting from individual task-instance outcome signals, it propagates error information backward through the evaluator, the agent's trajectory, and the context artifacts that shaped the agent's behavior, producing per-instance textual gradients.
\name{} then accumulates these gradients across instances by clustering, ranking, and filtering them, committing updates only when they reflect cross-instance consensus.
A convergence detector monitors the gradient signal itself, stopping optimization when the context has absorbed the generalizable information available from the training data, without requiring held-out validation data.
We evaluate \name{} on CyberGym, a benchmark of real-world vulnerability reproduction tasks. 
\rev{With fewer than 200 training examples, \name{} yields a 17-point self-improvement for GPT-5.4-mini (with Codex), enough to surpass the frontier GPT-5.4 baseline by 6 points, and an 18.5-point self-improvement for GPT-5.4 itself.
The resulting skills also transfer across models, highlighting that \name{}-optimized skills learn generalizable task knowledge rather than model-specific patterns.
Overall, \name{} enables agents to systematically improve themselves through context optimization and outperforms existing context optimization baselines.}
\end{abstract}


% \fi
\IEEEpeerreviewmaketitle

\section{Introduction}
\label{sec:intro}

LLM-based coding agents such as Codex~\cite{openai_codex_nodate}, Claude Code~\cite{anthropic_claude-code_nodate}, and others have become commonplace in software development, from everyday tasks to complex workflows.
These agents have also demonstrated strong software-security capabilities across the vulnerability lifecycle: discovering zero-days in production software~\cite{project-zero_big-sleep_nodate, liu_agentflow_2026,anthropic_mythos_nodate}, reproducing CVEs end-to-end across diverse languages~\cite{ullah_cve-genie_2026}, generating security patches at scale~\cite{google_codemender_nodate}.
To evaluate these capabilities systematically, the security community has developed a growing suite of benchmarks~\cite{wang_cybergym_2026, zhang_bountybench_2025, zhu_cvebench_2025, mei_arvo_2024}.
%
% The results are promising, but also show substantial room for improvement: for example, on CyberGym~\cite{wang_cybergym_2026}, the best agents in mid-2025 achieve only $\sim$20\% success at generating proof-of-concept (PoC) exploits for real-world vulnerabilities.

This raises a practical question: when an agent underperforms on a new task or domain, \emph{can it improve its own behavior using data from that domain, without human re-engineering or a stronger model?}
Improving an agent for a new task first requires deciding \emph{what} to optimize: the underlying LLM, the orchestration harness, or the agent context.
For security tasks, improving the underlying model is often computationally prohibitive or impractical as it requires expert traces, instrumented environments and reliable rewards, all of which are scarce in security domains~\cite{ullah_cve-genie_2026, risse_top-score_2025}.
Harness engineering faces another bottleneck: designing orchestration logic requires domain expertise and extensive manual effort~\cite{ullah_cve-genie_2026, li_patchpilot_2025}.
This leaves the agent's \emph{context} (its instructions, skills, and reference documents) as the most accessible component for task-specific improvement.

In modern coding agents, this context is organized as an \emph{agent home} (Figure~\ref{fig:agent-home}): a directory of instruction files, skill bundles, and sub-agent specifications, loaded automatically into the agent's context or retrieved on demand\cite{agent-skills_nodate}.
Developers continually refine this directory to improve task-specific performance.
However, like harness engineering, context engineering requires domain expertise, and manually designing context for every new task does not scale.
To reduce this burden, recent work has proposed automated context-optimization loops in which an agent runs on tasks, observes feedback from task outcomes, and uses an LLM to revise its context artifacts~\cite{zhang_ace_2025, ye_mce_2026, yuksekgonul_textgrad_2024, liu_agentflow_2026, lee_meta-harness_2026}.
This loop resembles classical \emph{gradient descent}: agent executions produce a \textit{loss} signal $\mathcal{L}$, such as a correct or incorrect outcome; an \textit{optimizer} LLM (the same underlying model as the agent) converts that signal into textual feedback, analogous to a \textit{gradient} $g$; and the resulting update modifies the context artifacts, the learnable \textit{parameters} $\theta$.

Existing context-optimization methods make this loop possible, but they remain poorly suited to long-running security tasks.
Their limitations fall along three axes: how they produce updates, how they validate those updates, and what kinds of supervision they can use.

First, existing methods typically run optimization in a \emph{single-shot} manner: they give all agent executions to the optimizer at once and ask it to infer how the context $\theta$ should be updated to improve outcomes.
In this setting, the optimizer never reasons about what the agent should have done independently of the existing context.
As a result, it cannot reliably distinguish whether the agent strayed from correct approach because of bad context, despite good context, or due to missing context altogether.
This also makes it difficult to calibrate the \emph{magnitude} of each update to $\theta$, which can lead to context collapse~\cite{zhang_ace_2025} or unnecessary complexity~\cite{li2026skillsbench}.

Second, existing methods often rely on held-out validation performance to distinguish generalizable updates from harmful noisy ones and decide when to stop.
This design is natural in ML optimization, where the dominant cost is computing gradients and validation on a large, representative set is comparatively cheap.
Agent context optimization has the opposite cost structure: producing gradient-like feedback requires only LLM inference over observed outcomes, whereas generating those outcomes requires executing the full agent pipeline in an instrumented environment.
Thus, every additional evaluation round carries disproportionate cost.
In practice, this forces methods to use small validation sets, which are unreliable indicators of actual task performance and generalization.
Validation-free methods avoid this cost, but by applying all updates without separating signal from noise and stopping after a fixed budget, they risk overfitting to the training tasks.

Third, security tasks often lack canonical ground truths.
Unlike domains such as mathematics, where problems have canonical solutions, vulnerability tasks admit many answers.
A vulnerability $v$ can be triggered by distinct PoCs and fixed by distinct patches; each valid solution $a$ belongs to a broad \emph{equivalence class} $\mathcal{E}_v$ defined by an underlying \emph{structural invariant} $\phi_v$: the root cause, trigger conditions, and failure mode of $v$.
Optimizing against one solution from this class risks shortcut learning~\cite{geirhos_shortcut_2020}.
Context optimization for security tasks must therefore often rely on sparse binary rewards, such as whether a generated PoC triggers the vulnerability, rather than dense supervision from canonical solutions.

We address these challenges with \name{}, an optimization algorithm that enables a security agent to improve its task context $\theta$ through textual gradient descent, without modifying the underlying model $\mathcal{M}$ or harness $\mathcal{H}$.
\name{} has four components. 
First, \emph{chain-rule gradient computation} propagates error backward from the loss $\mathcal{L}$, through evaluation scores $\mathbf{s}$ from the instrumented environment and the agent trajectory $\boldsymbol{\tau}$, to the context artifacts $\theta$.
This produces staged feedback that identifies where the failure originated.
Second, \emph{gradient accumulation} clusters, ranks, and filters per-task gradients $g$ by cross-task consensus.
It then annotates each surviving gradient against the current context as either \emph{hard} $g^{hard}$, indicating a gap not yet addressed, or \emph{soft} $g^{soft}$, indicating a pattern already covered but insufficiently enforced.
These annotations guide \emph{which} updates to apply and \emph{how aggressively}.
Third, a \emph{convergence detector} monitors the novelty and cross-task support of gradients across optimization steps.
When gradients become repetitive, noisy, or weakly supported, \name{} treats the optimization signal as exhausted and stops without held-out validation.
Finally, a \emph{vulnerability metadata supervisor} uses domain metadata $\mu_v$, such as vulnerability descriptions, patch commits, and security advisories, to guide gradient computation toward the structural invariant $\phi_v$ of the vulnerability: its root cause, trigger conditions, and failure mode, supporting optimization even when canonical ground truth is unavailable.

\begin{figure}[t]
    \centering
    \includegraphics[width=\linewidth]{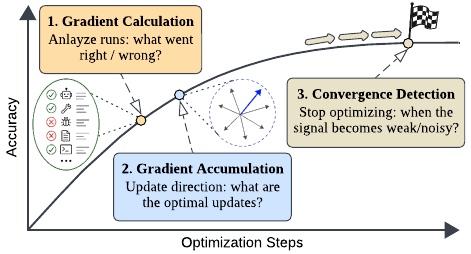}
    \caption{Optimization stages and recurring questions at each optimization step: How to analyze? Which direction to update towards and how much? Is the signal-to-noise ratio in the gradients strong enough to continue?}
    \label{fig:optimization-stages}
\end{figure}

We evaluate \name{} on CyberGym~\cite{wang_cybergym_2026}, a benchmark of real-world vulnerability-reproduction tasks.
With fewer than 200 training examples, \name{} enables GPT-5.4-mini with Codex to self-improve from 38\% to \rev{55\%} proof-of-concept generation success, without modifying the model or relying on a stronger teacher.
This \rev{17}-point gain enables the optimized GPT-5.4-mini agent to outperform the frontier model GPT-5.4 (baseline 49\%), despite GPT-5.4 being 3.5$\times$ more expensive per token.
Applied to GPT-5.4 itself, \name{} further improves its already strong baseline by 18.5-points, showing that context-based self-optimization scales from smaller agents to frontier models.
Moreover, optimized skills transfer across models: GPT-5.4's skill lifts GPT-5.4-mini to 65\%, and GPT-5.4-mini's skill lifts GPT-5.4 to 62\%, indicating that \name{} produces generalizable task knowledge rather than model-specific artifacts.
In this paper, we make three contributions:

\begin{enumerate}[leftmargin=*]
\item We introduce \name{}, a textual-gradient optimization algorithm for agent task context, including skills, instructions, and reference artifacts. \name{} decomposes optimization into chain-rule gradient computation, gradient accumulation, update calibration, and validation-free convergence detection, and we systematically evaluate these design choices through ablations.

\item We show how vulnerability metadata can serve as structured supervision to optimize context for security tasks. In the absence of canonical solutions, \name{} uses descriptions, advisories, and patch commits to guide gradients toward the structural invariant of each vulnerability: its root cause, trigger conditions, and failure mode.

\item We evaluate \name{} on CyberGym, a benchmark of real-world vulnerability-reproduction tasks. Using fewer than 200 training examples, \name{} self-improves agent skills for PoC generation, raising GPT-5.4-mini from 38\% to \rev{55\%} success and GPT-5.4 from 49\% to 67.5\%, without modifying the model, changing the harness, or relying on a stronger teacher.

\end{enumerate}

\section{Background and Related Work}

\subsection{Cybersecurity in the Era of LLM Agents}

\descr{LLM Agents in the Vulnerability Lifecycle.}
LLM agents are becoming capable across the full vulnerability lifecycle: discovery, reproduction, exploitation, and patching. DARPA's AI Cyber Challenge (AIxCC, 2023--2025)~\cite{zhang_aixcc-sok_2026} catalyzed this trend, challenging teams to build fully autonomous Cyber Reasoning Systems that discover and patch vulnerabilities in real-world open-source software, with finalist systems identifying 86\% of competition vulnerabilities and patching 68\% of those found~\cite{darpa_aixcc-results_nodate}. Industry efforts have followed: Google's Big Sleep discovered a zero-day in SQLite~\cite{project-zero_big-sleep_nodate}, DeepMind's CodeMender autonomously generated 72 security patches across open-source projects~\cite{google_codemender_nodate}, and Anthropic's Claude Mythos Preview uncovered a decades-old vulnerability in OpenBSD's TCP stack~\cite{anthropic_mythos_nodate}. In the academic community, AgentFlow discovered ten Chrome zero-days, including two Critical sandbox escapes~\cite{liu_agentflow_2026}; CVE-GENIE automates end-to-end CVE reproduction across diverse languages and vulnerability types~\cite{ullah_cve-genie_2026}, and multi-agent patching systems like PatchAgent decompose repair into localized fault analysis, patch synthesis, and validation~\cite{yu_patchagent_2025}.

\descr{Security Benchmarks.}
To standardize the assessment of these capabilities, the community has developed cybersecurity benchmarks spanning different stages of the vulnerability lifecycle. CTF-based benchmarks such as NYU CTF Bench and Cybench~\cite{shao_nyuctf_2025, zhang_cybench_2025} evaluate agents on curated challenges in controlled settings. Benchmarks grounded in real-world software, such as, CyberGym~\cite{wang_cybergym_2026} for vulnerability reproduction, ExploitGym~\cite{wang_exploitgym_2026} for exploitation, CVE-Bench~\cite{zhu_cvebench_2025} and BountyBench~\cite{zhang_bountybench_2025} for targeted vulnerability tasks, and SWE-Bench~\cite{jimenez_swebench_2024} and ARVO~\cite{mei_arvo_2024} for patch generation, better capture practical complexity of the security tasks. SEC-Bench~\cite{jing_secbench_2025} complements these benchmarks with knowledge-based evaluation through multiple-choice and short-answer questions. Evaluation approaches vary across these benchmarks, from deterministic sanitizer-based differential testing, to capture-the-flag (CTF) combined with LLM-based judging, to static knowledge assessment.

These benchmarks enable rigorous measurement of agent performance across the vulnerability lifecycle. The next question is \textbf{\emph{how to systematically improve agents' performance}}; a problem that, in security, differs from other domains, constraining the available optimization approaches.

% -------------------------------------------------------------
% -------------------------------------------------------------

\subsection{Adapting Agents to Security Tasks}
\label{subsec:adapting-agents}

\descr{General Levers for Improving Agents.} Improving an agent's performance on a new task can be approached through three levers: training better models, engineering better harnesses, or optimizing agent's context. We explain each of these levers below, highlighting the constraints that security tasks impose on their applicability.

\descrit{\ul{1) Model Training.}} Current approaches train software or security-specialized models either through supervised fine-tuning (SFT) on traces distilled from stronger models~\cite{nie_vulnllmr_2025, tang_copatcher_2025}, or through reinforcement learning (RL) in synthesized vulnerable environments~\cite{pan_swegym_2025, jain_r2e_2025, zhu_termigen_2026}. However, dependence of SFT on a stronger teacher imposes an inherent ceiling: it can transfer existing capabilities to smaller models but cannot push the frontier beyond what the teacher achieves~\cite{gudibande_false-promise_2023}. RL avoids this dependency but requires instrumented execution environments per vulnerability/task, which are expensive to build and limited in diversity. Both also require access to model weights, incurring the extensive computational cost of training and hosting the model.

\descrit{\ul{2) Agent Harness Engineering.}} The agent harness (orchestration code, tools, sub-agents, and coordination logic) wraps an LLM and can dramatically impact performance (e.g., Codex~\cite{openai_codex_nodate} and Claude Code~\cite{anthropic_claude-code_nodate}). Expert-designed harnesses have enabled capabilities that standalone models could not achieve: CVE-Genie's~\cite{ullah_cve-genie_2026} multi-agent pipeline enabled models to reproduce CVEs end-to-end across diverse languages, and PatchPilot's~\cite{li_patchpilot_2025} decomposition of patching into structured workflow improved both accuracy and stability. However, designing effective harnesses requires deep domain expertise, motivating automated approaches: Meta-Harness~\cite{lee_meta-harness_2026} uses a coding agent to iteratively propose and evaluate harness code, AgentFlow~\cite{liu_agentflow_2026} searches over multi-agent topologies for vulnerability discovery via a typed graph DSL, and OpenSage~\cite{li_opensage_2026} enables LLMs to self-generate agent structures and toolsets at runtime. While powerful, current automated harness search approaches require a capable coding agent as the search driver and iterate through unguided propose-and-evaluate loops, accepting or rejecting harness candidates based solely on whether they improve held-out performance~\cite{liu_agentflow_2026, lee_meta-harness_2026}, making the process both computationally expensive and reliant on trial and error rather than systematic optimization.

\descrit{\ul{3) Agent Context Engineering.}} The simplest form of context adaptation is in-context learning: where we provide an LLM representative few-shot examples relevant to the given task~\cite{brown_few-shot_2020}. However, modern coding agents operate with a richer structure. Agents like Codex, Claude Code, and many others, maintain an \emph{agent home} (Figure~\ref{fig:agent-home}), a directory containing a custom instructions file (\texttt{AGENTS.md} or \texttt{CLAUDE.md}) automatically loaded at every session, sub-agent specifications, and other agent-specific metadata. 
Early approaches to context customization relied on a single monolithic block of instructions, i.e., a prompt for standalone LLMs, or the main \texttt{AGENTS.md} / \texttt{CLAUDE.md} file for modern coding agents, but as instructions accumulated, these monolithic blocks became unwieldy and performance degraded~\cite{zhang_ace_2025}.
To address this, Anthropic introduced \emph{Agent Skills}~\cite{agent-skills_nodate} that distributes context across a structured multi-file bundle with progressive disclosure for a given task: one \texttt{SKILL.md} (loaded into agent's context on skill activation), references (loaded on demand), and executable scripts (loaded/ran on demand), as shown in Figure~\ref{fig:agent-home-skill-example}.

Similar to harness engineering, manually authoring context requires domain expertise and does not scale, motivating automated approaches such as TextGrad~\cite{yuksekgonul_textgrad_2024}, ACE~\cite{zhang_ace_2025}, MCE~\cite{ye_mce_2026}, and CoEvoSkills~\cite{zhang_coevoskills_2026} that use LLMs to iteratively generate and refine context artifacts from execution feedback, where the \emph{learnable parameter $\theta$} is no longer model weights or harness code but the textual context itself. 
While far more accessible than the other two levers, context optimization shares some of their limitations: current approaches also rely on a capable model to perform the analysis and optimize using an unguided propose-and-evaluate loop. It introduces further challenges unique to textual context as a learnable parameter: instructions are interpreted non-deterministically by the agent, context can grow unboundedly across iterations leading to degraded performance~\cite{zhang_ace_2025}, and compliance with instructions varies across tiers of the context hierarchy.

\descr{Vulnerability Metadata as Structured Supervision.}
As discussed in Section~\ref{sec:intro}, security tasks impose  an additional constraint on optimization beyond the  methodological limitations discussed above: vulnerabilities do not have a canonical ground truth.
A vulnerability $v$ can be triggered by many distinct inputs  and fixed by many distinct patches, each a member of a broad  \emph{solution equivalence class} $\mathcal{E}_v$ defined by an underlying  \emph{structural invariant} $\phi_v$ (the root cause, trigger  conditions, and failure mode):
\begin{equation}
\mathcal{E}_v = \{\, a \mid a \models \phi_v \,\}
\label{eq:equiv-class}
\end{equation}
Optimizing against any specific reference artifact  $a^* \in \mathcal{E}_v$ risks shortcut learning~\cite{geirhos_shortcut_2020} and rationalization~\cite{zelikman_star_2022} problems, rather than learning $\phi_v$ itself. 
However, $\phi_v$ is not directly observable. What \emph{is} observable is metadata routinely available in vulnerability databases and tracking systems:
\begin{equation}
\mu_v = (\textit{desc}_v,\; \textit{patch}_v,\; 
         \textit{trace}_v,\; \ldots)
\label{eq:metadata}
\end{equation}
Crucially, $\mu_v$ characterizes $\phi_v$ without prescribing a specific $a^* \in \mathcal{E}_v$: a patch reveals relevant code paths, a description identifies the root cause, and a sanitizer trace pinpoints the failure mode. 
Together, they enable an approximation:
\begin{equation}
\hat{\phi}_v = \textsc{Reconstruct}(\mu_v)
\label{eq:reconstruct}
\end{equation}
that describes \emph{what makes any solution valid} rather than what any particular solution looks like. 
This reconstructed invariant $\hat{\phi}_v$ can serve as structured supervision for optimization: an optimizer that reasons against $\hat{\phi}_v$ can self-assess the agent's approach using the agent's own model $\mathcal{M}$, without requiring a stronger teacher or canonical ground truth. 
This reconstructed invariant opens a path to self-supervised optimization, but \emph{how} to structure the optimization loop itself, from computing gradients to accumulating them to knowing when to stop, remains an open problem regardless of the supervision signal available.

% -------------------------------------------------------------
% -------------------------------------------------------------

\subsection{Automated Agent Optimization}

\newcommand{\cmark}{\textcolor{green!55!black}{\checkmark}}
\newcommand{\pmark}{\textcolor{orange!90!black}{(\checkmark)}}
\newcommand{\xmark}{\textcolor{gray}{--}}

\begin{table}[t]
\centering
\small
\setlength{\tabcolsep}{2pt}
\begin{threeparttable}
\caption{Optimization capabilities across agent improvement 
methods. \textbf{Param~$(\theta)$}: learnable parameter. \textbf{Grad}: structured gradient computation. 
\textbf{Accum}: cross-task gradient accumulation. \textbf{Conv}: signal-based 
convergence detection. 
\textbf{Meta}: use of domain metadata $\mu_v$ as structured 
supervision. 
\cmark~=~fully addressed, 
\pmark~=~partially addressed, 
\xmark~=~not addressed,
n/a~=~non relevant.}
\label{tab:opt-comparison}
\begin{tabularx}{\columnwidth}{@{}l@{\hspace{7pt}}lcccc@{}}
\toprule
\textbf{Method} & \textbf{Param $(\theta)$}
  & \textbf{Grad} & \textbf{Accum} & \textbf{Conv} & \textbf{Meta} \\
\midrule
% VulnLLM-R~\cite{nie_vulnllmr_2025}     & model weights & n/a & n/a & n/a & n/a \\
\tnote{$\dagger$}TextGrad~\cite{yuksekgonul_textgrad_2024}      & graph var/prompt & \cmark & \xmark & \xmark & \xmark \\
GEPA~\cite{agrawal_gepa_2025}          & prompt             & \xmark & \xmark & \xmark & \xmark \\
\tnote{$\dagger$}ACE~\cite{zhang_ace_2025}           & text playbook & \pmark & \cmark & \xmark & \pmark \\
MCE~\cite{ye_mce_2026}           & skills (context)   & \xmark & \pmark & \xmark & \xmark \\
CoEvoSkills~\cite{zhang_coevoskills_2026}   & skill package      & \xmark & \xmark & \xmark & \pmark \\
Meta-Harness~\cite{lee_meta-harness_2026}  & harness code       & \xmark & \xmark & \xmark & \xmark \\
AgentFlow~\cite{liu_agentflow_2026}     & harness topology   & \xmark & \xmark & \xmark & \pmark \\
\midrule
\textbf{\name{}} & agent skill & \cmark & \cmark & \cmark & \cmark \\
\bottomrule
\end{tabularx}
\begin{tablenotes}
\small
\item[$\dagger$] Designed for single-step LLM tasks, not agentic frameworks.
\end{tablenotes}
\end{threeparttable}
\end{table}

Across all three levers, and regardless of whether the supervision is artifact-based or metadata-guided, or whether we are optimizing harness code or textual context, the optimization loop follows the same structure: (1) run the agent on tasks (forward pass), (2) observe binary outcomes (loss), (3) analyze what went wrong (backward pass), and (4) update the learnable parameters (optimizer step). Meanwhile, the learnable parameters themselves span a spectrum: from prompt strings~\cite{khattab_dspy_2023, yuksekgonul_textgrad_2024, agrawal_gepa_2025} to persistent context and playbooks~\cite{suzgun_dynamic-cheatsheet_2026, zhang_ace_2025} to multi-file skill packages~\cite{ye_mce_2026, zhang_coevoskills_2026} to full harness code and multi-agent topologies~\cite{lee_meta-harness_2026, liu_agentflow_2026}. 
% Additionally, all approaches must respect structural constraints on their parameter space (e.g., AgentFlow~\cite{liu_agentflow_2026} type-checks its graph DSL, Meta-Harness validates that generated code compiles), a requirement we formalize as an \emph{optimization policy} in Section~\ref{sec:approach}. 
Despite this variation, all automated approaches face \emph{three methodological challenges} (Figure~\ref{fig:optimization-stages}).

\descr{1) Gradient Computation: How to Learn from Runs?} The dominant approach in literature is \emph{single-shot} analysis: an optimizer LLM reads the agent's results and proposes changes in a single pass~\cite{shinn_reflexion_2023, zhang_ace_2025, agrawal_gepa_2025}. TextGrad~\cite{yuksekgonul_textgrad_2024} introduces structure by textual feedback gradients and ACE~\cite{zhang_ace_2025} does it by separating the analysis and context update phases, but both only apply to single-step LLM pipelines (one input, one output, no tool calls), not multi-step agentic trajectories involving tool use, file access, and sub-agent coordination. Approaches that do handle agentic settings, such as MCE~\cite{ye_mce_2026}, Meta-Harness~\cite{lee_meta-harness_2026}, and AgentFlow~\cite{liu_agentflow_2026}, provide the optimizer with rich diagnostic access (full filesystem of agent runs' artifacts) but still analyze in a \emph{single-shot} setting. Across all these approaches, three gaps persist. (1) First, the analysis may conflate \emph{what went wrong in the execution} with \emph{which parameter caused it} in a single undifferentiated pass, potentially producing vague and undirected feedback. (2) Second, most exhibit a negativity bias, analyzing only failures while ignoring successful runs that carry equally important signal about which instructions and behaviors to preserve. (3) In security, an additional gap arises: these approaches depend on either ground-truth labels/artifacts or a strong LLM to assess quality. None allow systematically leveraging the domain/vulnerability metadata $\mu_v$ (such as, patches, vulnerability descriptions, stack traces, etc.) to self-evaluate instead of depending on a teacher model.

\descr{2) Gradient Accumulation: How to Apply Learnings?} The most common approach among related works is to apply all feedback at once: the optimizer reads the full analysis and rewrites the learnable parameter in a single step~\cite{yuksekgonul_textgrad_2024, shinn_reflexion_2023}. When feedback comes from a batch of tasks, individual signals may be noisy, contradictory, or overwhelming, yet they are consumed as a monolithic input. Some approaches introduce structure: ACE~\cite{zhang_ace_2025} uses incremental delta updates with deduplication to control context bloat, and GEPA~\cite{agrawal_gepa_2025} maintains a Pareto frontier of diverse candidates, but again they both do not apply to multi-step agentic trajectories. Moreover, the majority of approaches gate updates through a held-out validation set, accepting changes only if they improve validation performance~\cite{yuksekgonul_textgrad_2024, agrawal_gepa_2025, lee_meta-harness_2026, liu_agentflow_2026}. Validation gating is expensive (each checkpoint requires full evaluation), noisy (small sets may reject changes that would generalize, see Section~\ref{subsec:convergence}), and makes the process trial-and-error rather than directed systematic optimization. None of the current approaches systematically prioritize \emph{which} changes to apply based on cross-task evidence. Moreover, even with prioritization and filtering, no existing approach distinguishes whether a surviving gradient targets a \emph{gap} in the current parameter $\theta$ (a pattern the context has never addressed) or a \emph{compliance failure} (a pattern already covered but insufficiently emphasized or poorly phrased), leaving the optimizer unable to calibrate the \emph{magnitude} of its edits, potentially leading to optimization collapse~\cite{zhang_ace_2025}. Furthermore, these approaches are sensitive to optimizer capability: stronger models may handle bulk feedback gracefully and make reasonable update decisions, but weaker models cannot, making the optimization dependent on model strength rather than algorithmic design. 
On top of this, in security, high-quality datasets are scarce and each evaluation requires running the full agent pipeline in an instrumented environment, compounding costs in terms of time and money. Even if we run the evaluation after each step, the held-out validation datasets still do not provide a reliable signal~\S\ref{subsec:convergence}.

\descr{3) Convergence Detection: When to Stop Optimizing?} Most approaches run the optimization loop for a fixed number of iterations~\cite{agrawal_gepa_2025, lee_meta-harness_2026} or until a compute budget is exhausted~\cite{liu_agentflow_2026}. Unlike neural network training, where the backward pass dominates cost and forward/evaluation pass is cheap, inference-based textual optimization inverts this: computing textual gradients involves a few LLM calls, while each forward pass and evaluation requires full agent execution, making any stopping criterion that depends on additional forward passes inherently expensive. In security tasks (e.g., CyberGym), where each forward pass can take upwards of 30 minutes per task in an instrumented environment, and several dollars per task, these costs are particularly prohibitive. Beyond their cost, validation runs also introduce noise because validation depends on the set size and representativeness (see Section~\ref{subsec:convergence}). No existing approach systematically monitors the optimization signal itself to detect \emph{convergence}: whether proposed changes across successive rounds are becoming weak, noisy, or repetitive, which would indicate that the optimization has extracted the generalizable information available from the training data without requiring any additional forward passes.

\begin{tcolorbox}[colback=gray!5, colframe=black!70,
  title=\textbf{Research Questions},
  boxrule=0.5pt, left=4pt, right=4pt, top=3pt, bottom=3pt]
\small
\begin{enumerate}[leftmargin=*, topsep=2pt, itemsep=2pt]
    \item \textbf{Self-Supervision:} How to leverage vulnerability metadata $\mu_v$ for self-assessment, without requiring a stronger optimizer model or canonical ground truth?
    \item \textbf{Gradient Computation:} How to decompose the backward pass for multi-step agentic trajectories, separating behavioral analysis from context attribution?
    \item \textbf{Gradient Accumulation:} How to identify the most generalizable update direction from noisy per-task gradients, and calibrate how aggressively to apply each update?
    \item \textbf{Convergence Detection:} How to detect that the optimization has extracted the generalizable signal available from the training data and what remains is noise?
\end{enumerate}
\end{tcolorbox}

\section{\name}

\label{sec:approach}

\begin{figure}[t]
    \centering
    \includegraphics[width=0.95\linewidth]{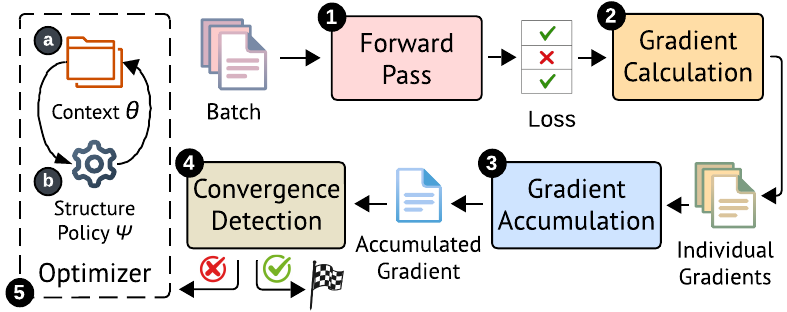}
    \caption{\name{} pipeline overview}
    \label{fig:pipeline}
\end{figure}

\name{} is an optimization algorithm that improves a frozen agent's context $\theta$ for a given task through \emph{textual gradient descent}, without modifying the underlying model $\mathcal{M}$ or harness $\mathcal{H}$.
The algorithm operates in discrete steps over sequential batches of training tasks (Figure~\ref{fig:pipeline}). 
At each step, the agent executes on a batch using the current context, producing binary pass/fail outcomes (\circled{1}~forward pass). 
A chain-rule-inspired backward pass then decomposes error attribution through three successive stages: evaluation scores, agent trajectory, and context artifacts; these produce per-task textual gradients that identify \emph{which} context deficiencies caused \emph{which} behavioral failures (\circled{2}~gradient computation). 
These per-task gradients are then clustered, ranked, and filtered by cross-task consensus, so that only patterns with sufficient evidential support survive, and each surviving gradient is annotated against the current context as either a \emph{hard} gradient (a gap not yet addressed) or a \emph{soft} gradient (a pattern already present but insufficiently effective), calibrating both the direction {\bf and} magnitude of the optimizer's edits (\circled{3}~gradient accumulation). 
Finally, a convergence detector monitors the novelty and signal-to-noise ratio of accumulated gradients across steps: as the gradients become repetitive or lose cross-task support, the optimization signal is exhausted and the process terminates, without requiring held-out validation (\circled{4}~convergence detection). 
The optimizer (\circled{5}) applies the accumulated, annotated gradients to \circled{a} context $(\theta)$, guided by a \circled{b} \emph{structure policy}: a \emph{document} that describes the valid syntactical format, layout, and constraints of the context $\theta$ for the given agent (e.g., YAML format for SKILL.md, Figure~\ref{fig:skill-example}), so that the optimizer knows \emph{what} it can edit and \emph{how} the context must be syntactically structured. 
The structure policy also provides a \emph{structure checker} to the LLM that is can use to validate its edits against the given policy to ensure the resulting context remains \emph{syntactically} loadable by the agent.

Rather than presenting \name{} as a monolithic system, we describe it through the investigative process. 
We first establish the investigation methodology (\S\ref{subsec:investigation-methodology}), and then present gradient computation (\S\ref{subsec:gradient-computation}), accumulation (\S\ref{subsec:gradient-accumulation}), and convergence detection (\S\ref{subsec:convergence}) as successive contributions, each validated by targeted ablation.

\subsection{Investigation Methodology}
\label{subsec:investigation-methodology}

\descr{Security Task (PoC Generation).} We evaluate on CyberGym~\cite{wang_cybergym_2026}, a benchmark of real-world vulnerability reproduction tasks. CyberGym is well-suited for our investigation for three reasons. First, the task presents real-world complexity: agents must reason about diverse codebases, vulnerability types, and build systems, with individual tasks taking up to 30 minutes of agent execution (consistent with CyberGym). Second, evaluation is fully deterministic: sanitizer output and differential testing across program versions verify whether a PoC triggers the target vulnerability, yielding a clean binary loss signal with no LLM-judge noise contaminating the optimization loop. Third, CyberGym's tasks are sourced from OSS-Fuzz, so each vulnerability ships with metadata: the OSS-Fuzz bug report, the developer patch commit, and the sanitizer stack trace, that we use as the structured supervision signal ($\mu_v$) for gradient computation.

\descr{Dataset Selection.} CyberGym contains 1{,}507 instances, but a single evaluation pass costs approximately \$3{,}000 in API credits~\cite{wang_cybergym_2026}; for an optimization study requiring repeated forward passes across multiple ablation configurations, this cost is prohibitive. We design our investigations in this section around a batch size of 16, and randomly sample 224 training tasks (${\sim}$15\% of CyberGym), yielding 14 optimization steps per epoch. We also sample 40 tasks (${\sim}$3\%) for validation and 80 (${\sim}$5\%) for test, for a total of 344 tasks (${\sim}$23\%). This scale is consistent with prior work that use 20--25\% subsets of CyberGym due to cost~\cite{wang_cybergym_2026, li_opensage_2026}. Although \name{}'s convergence detector does not use the validation set for stopping decisions (Section~\ref{subsec:convergence}), we retain it to track training dynamics in our ablations and to enable fair comparison with baselines that rely on validation-gated checkpoint selection.

\begin{figure}[h]
\centering
\subfloat[Agent home directory structure (\texttt{\~{}/.codex/}), with annotations showing when each component is loaded into the agent's context.]{\includegraphics[width=\columnwidth]{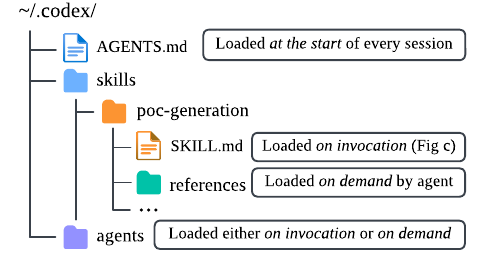}%
\label{fig:agent-home}}
\hfill
\subfloat[Structure of a \texttt{SKILL.md} file, illustrated with a \emph{poc-generation} skill targeting CyberGym's vulnerability reproduction task.]{\includegraphics[width=\columnwidth]{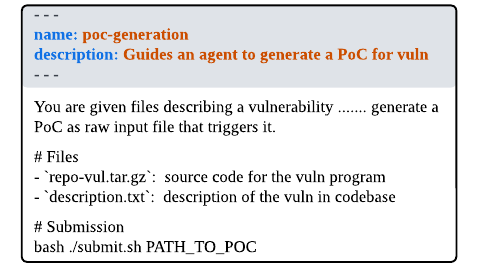}%
\label{fig:skill-example}}
\hfill
\subfloat[Skill invocation at the command line; the agent loads the corresponding \texttt{SKILL.md} and begins execution.]{\includegraphics[width=\columnwidth]{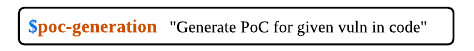}%
\label{fig:invoke-example}}
\caption{Agent home and skills. The agent home directory~(a) structures context into a persistent instructions file, modular skills, and sub-agents, each loaded at different stages. A skill bundles a \texttt{SKILL.md} file with references and scripts~(b), and is activated by name at invocation time~(c).}
\label{fig:agent-home-skill-example}
\end{figure}

\descr{Skill as Learnable Parameter $(\theta)$.} We scope our learnable parameters to a \emph{single} \textbf{agent skill}~\cite{agent-skills_nodate}, which we name \texttt{\ul{poc-generation}}: a \texttt{SKILL.md} file loaded on skill activation (Figure~\ref{fig:invoke-example}) and associated \texttt{reference} documents loaded on demand during task execution. We exclude sub-agent specifications and executable scripts from the learnable parameter space because isolating the simplest textual parameter space lets us study the optimization methodology without confounding it with the complexity of optimizing executable artifacts. We discuss extension to broader parameter spaces in Section~\ref{sec:discussion}. For this investigation, at the start of each optimization run, we initialize skill with CyberGym's base prompt (as shown in Figure~\ref{fig:skill-example}).

\descr{Agent, LLM, and Structure Policy.} We use OpenAI Codex as the frozen agent harness with GPT-5.4-mini as the underlying model. Codex's built-in sub-agents (explore, worker, default) are part of the frozen harness and are not modified. The optimizer is also Codex using the same model, meaning we are not relying on a stronger model for optimization, allowing us to study \emph{whether a model can self-improve its own context}. For this study, our \emph{structure policy} is derived from Codex's official documentation: it describes the skill format (YAML frontmatter, file layout, per-file token limits), how reference documents must be cited from the main \texttt{SKILL.md}, and the available built-in runtime tools and sub-agent Codex offers that the skill can leverage. The \emph{structure checker} enforces these constraints after every optimizer step without any semantic evaluation of the skill's content.

\descr{Training Loop.} Optimization proceeds in discrete steps over sequential batches of training tasks (Figure~\ref{fig:pipeline}). At each step, the agent runs on a batch of tasks using the current \texttt{poc-generation} skill (forward pass), each task is scored via execution-based evaluation producing a binary pass/fail (loss), and the optimizer then receives either raw results or pre-computed textual gradients, depending on the configuration under study, edits the skill (backward pass and optimizer step). The optimizer persists across steps within a training run, retaining memory of all prior edits, their motivations, and their effects on subsequent batches. We do not ablate this design choice of persistent history, as prior work has already demonstrated the importance of maintaining optimization history across iterations~\cite{lee_meta-harness_2026, liu_agentflow_2026}. Our full configuration for this investigation is as follows.

\begin{tcolorbox}[colback=gray!5, colframe=black!70,
  title=\textbf{Experimental Configuration},
  boxrule=0.5pt, left=3pt, right=3pt, top=3pt, bottom=3pt, label={box:config}]
\small
\begin{tabular}{@{}p{0.28\columnwidth}p{0.62\columnwidth}@{}}
\textbf{Harness $\mathcal{H}$} & OpenAI Codex (frozen) \\
\textbf{Model $\mathcal{M}$} & GPT-5.4-mini (agent \& optimizer) \\
\textbf{Param ($\theta$)} & \texttt{poc-generation} skill (text only)  \\
\textbf{Benchmark} & CyberGym (vuln descr. $\xrightarrow{}$ PoC) \\
\textbf{Dataset} & 224 train / 40 val / 80 test \\
\textbf{Batch Size $m$} & 16 tasks (14 steps in 1 epoch) \\
\textbf{Time Limit} & 30 min per task \\
\textbf{Evaluation} & Sanitizer \& differential test (pass@1) \\
\textbf{Baseline $\theta_0$} & CyberGym base prompt (20\% val) \\
\end{tabular}
\end{tcolorbox}

\begin{figure}[t]
    \centering
    \includegraphics[width=0.95\linewidth]{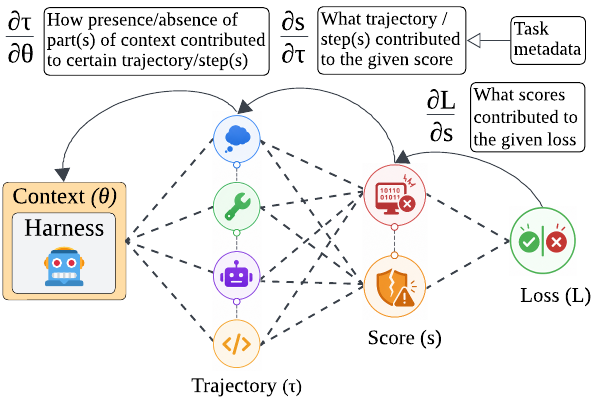}
    \caption{Gradient Computation: chain-rule decomposition of backward pass for an agentic forward pass.}
    \label{fig:grad-comp}
\end{figure}

\begin{figure*}[t]
\centering
\small
\setlength{\fboxsep}{6pt}

% ── Stage 1 ──────────────────────────────────────────────────────
\noindent\fbox{\parbox{0.96\textwidth}{%
\textbf{Stage~1: Score Analysis}
\hfill $\partial \mathcal{L} / \partial \mathbf{s}$\\[4pt]
%Task: arvo\_3956 \hfill Overall: \ding{55}~(incorrect)\\[2pt]
\begin{tabular}{llccc}
\toprule
Task & PoC ID & Crashed Binary & Triggered Vuln & Overall \\
\midrule
arvo\_3956 & \texttt{poc-0112367f...} & \ding{55} & \ding{55} & \textbf{Incorrect} \\
\bottomrule
\end{tabular}\\[2pt]
%{\footnotesize Summary: 1 submitted, 0 crashed binary, 0 triggered vuln}
}}

\vspace{6pt}

% ── Stage 2 ──────────────────────────────────────────────────────
\noindent\fbox{\parbox{0.96\textwidth}{%
\textbf{Stage~2: Trajectory Analysis}
\hfill $\partial \mathbf{s} / \partial \boldsymbol{\tau}$
\quad\emph{(skill-blind)}\\[4pt]
\textbf{Score Analysis:} [\ldots] The agent's final \texttt{poc.txt}
was effectively a copy of curl's \texttt{tests/data/test1800}, which
is a reference HTTP/2 upgrade regression testcase, not a proof that
the OOM bug is reachable. [\ldots]\\[3pt]
\textbf{Trajectory Strengths:} [\ldots] It correctly found the
relevant regression test files (\texttt{test1800} and
\texttt{test1801}) and read the HTTP/2 implementation around the
upgrade/switch path. [\ldots]\\[3pt]
\textbf{Trajectory Weaknesses:} It overfit to the reference testcase
instead of deriving a fault-triggering input from the vulnerability
description. [\ldots] It conflated ``matching the known test layout''
with ``triggering the bug.'' [\ldots]\\[3pt]
\textbf{Trajectory Improvement Suggestions:} Treat reference tests as
scaffolding only. Use them to understand protocol shape, then mutate
the specific bytes or sequence that influence the vulnerable code
path. [\ldots]
}}

\vspace{6pt}

% ── Stage 3 ──────────────────────────────────────────────────────
\noindent\fbox{\parbox{0.96\textwidth}{%
\textbf{Stage~3: Orchestration Analysis}
\hfill $\partial \boldsymbol{\tau} / \partial \theta_{\text{skill}}$
\quad\emph{(skill-aware)}\\[4pt]
\textbf{Agent's Usage of Context:} The agent used the
\texttt{poc-generation} skill in the intended broad sense: inspect
\texttt{description.txt}, inspect the vulnerable source, and submit
through \texttt{submit.sh}. [\ldots]\\[3pt]
\textbf{Strengths:} [\ldots] The run did not wander into unrelated
code; it stayed centered on the described HTTP/2 upgrade flow [\ldots]\\[3pt]
\textbf{Weaknesses:} \texttt{SKILL.md} tells the agent to generate
a PoC, but it does not tell it how to distinguish a reference
regression testcase from a vulnerability-triggering input. That gap
maps directly to the observed overfitting to \texttt{test1800}.
[\ldots]\\[3pt]
\textbf{Improvements:} Add negative instructions directly in
\texttt{SKILL.md}: do not stop at a known regression testcase, do
not treat protocol-format parity as success [\ldots]
}}

\vspace{4pt}
% \caption{Three-stage gradient for task \texttt{arvo\_3956}, showing
% the output format of each stage. Stage~2 contains the behavioral
% grounding, \emph{why} regression tests cannot serve as
% PoCs, under `Trajectory Weaknesses' and `Improvement
% Suggestions'. Stage~3 contains the skill attribution and proposed
% edit under `Weaknesses' and `Improvements'.}

\caption{Outputs from each gradient-computation stage for task \texttt{arvo\_3956}. Stage~2 grounds why regression tests cannot serve as PoCs under `Trajectory Weaknesses' / `Improvement Suggestions'. Stage~3 shows skill attribution and proposed edits under `Weaknesses' / `Improvements'.}

\label{fig:three-stage-gradient}
\end{figure*}

\subsection{Gradient Computation}
\label{subsec:gradient-computation}

\descr{The Agentic Forward Pass.} In a single-step setting, the forward pass maps an input directly to an output (i.e., the object of evaluation). In an agentic setting, the forward pass is a multi-stage process (Figure~\ref{fig:grad-comp}). The agent, guided by the skill $\theta$, produces a \emph{trajectory} $\mathcal{T}$: a sequence of reasoning steps, tool calls, and sub-agent interactions, often spanning hundreds of actions. The trajectory submits \emph{artifacts} (e.g., a raw PoC files), which are then evaluated in an execution environment to produce \emph{scores} $\mathbf{s}$: a vector of per-criterion outcomes (e.g., did the PoC crash the vulnerable version? did the sanitizer trigger and the correct vulnerability was triggered?). The final binary \emph{loss} $\mathcal{L}$ is an aggregation of these scores. This layered structure means the backward pass cannot simply ask ``how should the skill change to improve the output,'' as in single-step settings~\cite{yuksekgonul_textgrad_2024}. It must reason backward through scores, trajectory, and artifacts to understand which agent behaviors led to the outcome.

\descr{The Backward Pass Problem.} The majority approach to LLM-based gradient computation is \emph{single-shot}: the optimizer receives the task outcomes and the current learnable parameters and proposes improvements in a single pass~\cite{ye_mce_2026, zhang_ace_2025, liu_agentflow_2026, lee_meta-harness_2026}. This conflates several distinct reasoning steps: (a) inferring what the optimal approach \emph{should} have been, (b) determining what the agent actually did relative to that optimum, (c) identifying the skill's role in guiding those behaviors, and (d) proposing edits that would have steered the agent differently. A sufficiently strong model may handle this implicitly, but when the optimizer is not proficient at the target task, single-pass reasoning becomes unreliable: the optimizer lacks an independent basis for %judgment and instead anchors on surface-level patterns in the trajectory.

As discussed in Section~\ref{subsec:adapting-agents}, this problem is compounded in security tasks by the absence of canonical ground truth. Before the optimizer can assess the agent's trajectory, it must first reconstruct the structural invariant $\hat{\phi}_v$ from the available metadata $\mu_v$ (Eq.~\ref{eq:reconstruct}): understanding the root cause, trigger conditions, and failure mode of the vulnerability \emph{independently} of what the agent attempted. Only with this reconstructed invariant can the optimizer meaningfully evaluate whether the agent's approach was sound and whether the skill should have guided it differently, rather than judging against a specific artifact, $a^* \in \mathcal{E}_v$, that risks the shortcut learning problem (Eq.~\ref{eq:equiv-class}). A single-shot optimizer that skips this independent reconstruction of $\hat{\phi}_v$ risks proposing skill edits that address symptoms in the trajectory rather than the %underlying gap in the agent's approach.

\descr{Chain-Rule Decomposition.} We decompose the backward pass using a chain-rule analogy (Figure~\ref{fig:grad-comp}):
\begin{equation}
\frac{\partial \mathcal{L}}{\partial \theta_{\text{skill}}} = \frac{\partial \mathcal{L}}{\partial \mathbf{s}} \cdot \frac{\partial \mathbf{s}}{\partial \boldsymbol{\tau}} \cdot \frac{\partial \boldsymbol{\tau}}{\partial \theta_{\text{skill}}}
\label{eq:chain-rule}
\end{equation}
where $\mathcal{L}$ is the binary loss, $\mathbf{s}$ the evaluation scores, $\boldsymbol{\tau}$ the agent trajectory, and $\theta_{\text{skill}}$ the learnable skill parameters. Each factor is computed by a separate analysis stage with distinct information access and the analysis from the previous stages. We illustrate each stage with task \texttt{arvo\_3956} (Figure~\ref{fig:three-stage-gradient}) where the agent had to produce a PoC triggering an out-of-memory vulnerability in curl's HTTP/2 handling.

\descrit{\ul{Stage~1}} ($\partial \mathcal{L} / \partial \mathbf{s}$): decompose the binary loss into its constituent score criteria to identify \emph{where} in the evaluation pipeline the task succeeded or failed. 
In CyberGym, the loss decomposes into: did the agent produce a PoC file? Did the PoC crash the pre-patch binary? Did the sanitizer detect the target vulnerability? In \texttt{arvo\_3956}, the agent submitted a single PoC that neither crashed the binary nor triggered the vulnerability (Figure~\ref{fig:three-stage-gradient}, Stage~1), pinpointing the failure at the earliest evaluation criterion.

\descrit{\ul{Stage~2}} ($\partial \mathbf{s} / \partial  \boldsymbol{\tau}$, \emph{skill-blind}): given the score decomposition, trajectory, and task metadata $\mu_v$,  analyze which agent behaviors led to those specific outcomes. 
In \texttt{arvo\_3956}, Stage~2 (Figure~\ref{fig:three-stage-gradient}) reveals that the submitted PoC was effectively a copy of curl's regression test \texttt{test1800}, a reference for the \emph{fix} in curl, not a vulnerability trigger. While the agent correctly located relevant test files and read the HTTP/2 upgrade implementation, it conflated matching the known test layout with triggering the bug. This stage recommends treating reference tests as scaffolding only: use them to understand protocol shape, then mutate the fields influencing the vulnerable code path.
Critically, this stage does \emph{not} see the skill~$\theta$, preventing the analysis from being anchored on what the skill instructed rather than what the agent actually did. 
Instead, the optimizer reconstructs $\hat{\phi}_v$ from $\mu_v$ (Eq.~\ref{eq:reconstruct}) and evaluates the trajectory against this structural invariant rather than against any specific artifact $a^* \in \mathcal{E}_v$, providing the metadata-guided self-supervision as follows:
\begin{equation}
\frac{\partial \mathbf{s}}{\partial \boldsymbol{\tau}} \bigg|_{\hat{\phi}_v}
\quad\text{where}\quad
\hat{\phi}_v = \textsc{Reconstruct}(\mu_v)
\label{eq:metadata-supervision}
\end{equation}

\descrit{\ul{Stage~3}} ($\partial \boldsymbol{\tau} / \partial \theta_{\text{skill}}$, \emph{skill-aware}): given the Stage~2 behavioral analysis \emph{and} the current skill, identify which specific skill instructions (presence or absence) caused the problematic behaviors and propose targeted changes. 
In \texttt{arvo\_3956}, Stage~3 traces this behavioral gap to a missing negative instruction: \texttt{SKILL.md} directs the agent to generate a PoC but never warns against treating regression tests as vulnerability triggers. The proposed edit adds explicit guidance: do not submit existing regression tests as PoCs, and do not treat protocol-format parity as success.

\descr{Experimental Ablation.} To validate that chain-rule decomposition produces better gradients than previous approaches, we design four optimizer variants along a spectrum:

\begin{itemize}[leftmargin=*, topsep=4pt, itemsep=2pt]
\item \emph{D1 (Single-Shot):} This mirrors the approach in prior work, such as MCE~\cite{ye_mce_2026}, Meta-Harness~\cite{lee_meta-harness_2026}, and Agentflow~\cite{liu_agentflow_2026}, where the optimizer agent receives all batch results and $\theta$, and proposes changes in a single pass.
\item \emph{D2 (Per-Task):} Each task gets an independent single-stage analysis that directly computes $\partial \mathcal{L} / \partial \theta_{\text{skill}}$ in one pass. The main optimizer reads all per-task analyses and applies updates. This mirrors how optimization approaches like ACE~\cite{zhang_ace_2025} and CoEvoSkills~\cite{zhang_coevoskills_2026} operate.
\item \emph{D3 (Chain-Last):} Full three-stage chain-rule decomposition per task, but the main optimizer sees only the Stage~3 output ($\partial \boldsymbol{\tau} / \partial \theta_{\text{skill}}$) for all tasks.
\item \emph{D4 (Chain-All):} Same chain-rule decomposition, but the main optimizer sees all three stage outputs for all tasks. Intermediate gradients (score decomposition, behavioral analysis) are provided alongside Stage~3 analysis. This is used as the gradient computation component in \name.
\end{itemize}

\noindent Each adjacent comparison isolates a specific hypothesis. D1~vs.~D2 tests whether per-task decomposition improves over batch-level single-shot analysis. D2~vs.~D3 tests whether chain-rule decomposition through the trajectory produces better gradients than a direct single-stage mapping from loss to skill. D3~vs.~D4 tests whether preserving intermediate gradients provides additional value to the optimizer beyond the final output of the chain analysis.

\begin{figure}[t]
\centering
\subfloat[D1 (Single-Shot)]{\includegraphics[width=0.495\columnwidth]{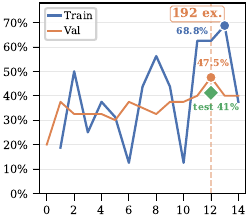}%
\label{fig:d1}}
\hfill
\subfloat[D2 (Per-Task)]{\includegraphics[width=0.495\columnwidth]{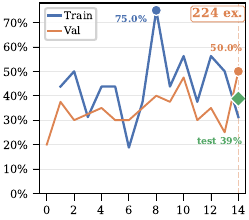}%
\label{fig:d2}}\\
\subfloat[D3 (Chain-Last)]{\includegraphics[width=0.495\columnwidth]{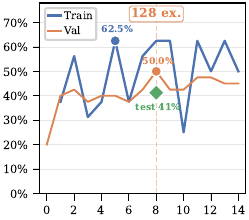}%
\label{fig:d3}}
\hfill
\subfloat[D4 (Chain-All)]{\includegraphics[width=0.495\columnwidth]{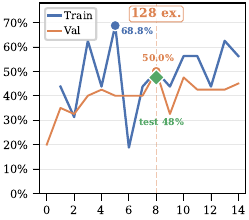}%
\label{fig:d4}}
\caption{Train vs.\ validation accuracy across gradient decomposition variants (x-axis: optimizer steps; y-axis: accuracy). Each plot shows training (blue) and validation
accuracy (orange) for each step. Green diamond marks test accuracy at the best validation checkpoint. Dashed lines indicate the number of training examples used to reach the best checkpoint.}
\label{fig:gradient-ablation}
\end{figure}

\descr{Results.} We run all four designs on identical CyberGym training batches using the config above (Figure~\ref{fig:gradient-ablation}).
% We further categorize the results as follows.

\descrit{Sample Efficiency.} D3 and D4 both reach their best validation accuracy (50\%) at step~8 using 128 training examples. D2 requires all 224 examples (step~14) to match 50\%, and D1 never does, peaking at 47.5\% at step~12. Chain-rule decomposition thus provides a 1.75$\times$ improvement in sample efficiency over per-task analysis for the same validation performance: structuring the backward pass along the causal chain extracts more learning signal per training example.

\descrit{Generalization.} Despite D2, D3, and D4 all reaching 50\% validation accuracy, they diverge sharply on the held-out test set: D4 achieves 48\%, while D3 and D1 both reach 41\%, and D2 falls to 39\%. Two observations stand out. First, D2 exhibits the worst test generalization despite matching the best validation score, suggesting that per-task decomposition without chain-rule structure produces gradients that overfit to training-specific patterns. Second, D3 and D4 use the same three-stage chain, and Stage~3 already incorporates the reasoning from earlier stages in its output. Yet D4 outperforms D3 by 7 points. The difference is that D4 exposes the full causal chain to the main optimizer: which score criteria failed, what behavioral mistakes caused those failures, and how the skill contributed. 
This allows the optimizer to trace each recommendation back to concrete behavioral evidence and write grounded, transferable guidance. For instance, in \texttt{arvo\_3956}, the full chain produces: ``Do not submit existing regression tests as PoCs. They exercise the fixed behavior, not the vulnerability. Use them to understand the protocol shape, then mutate the fields that influence the vulnerable code path.'' This behavior-guided instruction is missed without intermediate stages.

\begin{tcolorbox}[colback=blue!3, colframe=black!50,
  title=\textbf{RQ1,2: Self-Supervision \& Gradient Computation},
  boxrule=0.5pt, left=4pt, right=4pt, top=3pt, bottom=3pt]
\small
Three-stage chain-rule decomposition yields $\mathbf{1.75\times}$
better sample efficiency than single-stage per-task analysis, further adding a \textbf{7-point test accuracy gain}. 
The key mechanism is the \textbf{skill-blind} middle
stage: by reconstructing the structural invariant ($\psi_v$) from vulnerability
metadata ($\mu_v$) \emph{before} seeing the skill, the optimizer produces
edits grounded in \textbf{behavioral evidence} rather than
surface-level pattern matching.
\end{tcolorbox}

\subsection{Gradient Accumulation}
\label{subsec:gradient-accumulation}

\descr{Motivation.} While D4 produces the highest-quality per-task gradients of the four designs, analyzing its optimization trajectory reveals systematic pathologies (Figure~\ref{fig:gradient-ablation}) that prevent these gradients from translating into stable, generalizable skill improvements. We identify two problems: \emph{direction}: where the optimizer cannot determine \emph{which} gradients to trust, and \emph{magnitude}: where even correctly identified gradients carry no signal about \emph{how aggressively} to edit.

\descrit{(1) \ul{Direction: Which gradients to trust?}} When per-task gradients are provided directly to the optimizer without aggregation or filtering, four pathologies emerge:

\begin{itemize}[leftmargin=*, topsep=4pt, itemsep=2pt]
\item \emph{Negativity bias:} The optimizer inspected only failing tasks in 13 of 14 steps, never observing what the current skill does well. Successful runs carry equally important signal about which instructions and behaviors to \emph{preserve}, yet this signal is systematically ignored.

\item \emph{Strength erosion:} In 57\% of steps, at least one adopted change contradicted a demonstrated strength. The three worst accuracy regressions coincide with the highest strength-erosion counts: the optimizer overwrites effective guidance because it never identified it as effective.

\item \emph{Low-support noise:} Changes were adopted with support from as few as 1 out of 12 failing tasks, injecting task-specific heuristics into a skill, affecting generalizability across diverse vulnerability types.

\item \emph{Contradiction adoption:} 78\% of contradictory gradient pairs (e.g., ``broaden search'' vs.\ ``lock onto one path'') had both directions adopted, producing internally inconsistent guidance that leaves the agent without a clear directive.

\end{itemize}

\noindent These flaws are analogous to stochastic gradient descent with batch size~1 and an excessive learning rate: individual gradients are too noisy and contradictory to reliably and generalizable optimization.

\descrit{(2) \ul{Magnitude: How aggressively to edit?}} Even if the direction problems were resolved and only well-supported, non-contradictory gradients survived, a second pathology remains: the optimizer receives all surviving improvements as undifferentiated items, with no signal about the \emph{nature} of each edit.
Consider two improvement suggestions: ``agents should analyze the data flow path of the vulnerable function before crafting a PoC'' and ``agents should verify their PoC triggers the correct sanitizer.'' 
The first may describe a \emph{gap}: the skill~$\theta$ contains no guidance on data flow analysis whatsoever. 
The second may describe a \emph{compliance failure}: the skill already instructs sanitizer verification, but the agent ignored or misapplied it. 
These require fundamentally different edits: the former may require new content, while the latter may require only reinforcement or rephrasing of existing content.
Without this distinction, the optimizer treats both identically: it re-adds existing guidance when it should reinforce, and tentatively rephrases when it should introduce new content. 
This miscalibration is also the primary driver of \emph{unbounded complexity growth}: because the optimizer cannot recognize that a pattern is already covered in~$\theta$, it appends redundant instructions rather than strengthening existing ones, growing the skill from 122 words and 1 file to 2{,}612 words and 13 files over 14 steps (see Figure~\ref{fig:skill-growth}). 
The bloat reflects not too many updates, but it is a symptom of the optimizer lacking information to make \emph{precise} updates.

\descr{Solution.} We introduce an accumulation stage
that sits between \circled{2} per-task gradient computation (D4's output) and the \circled{5} main optimizer (Figure~\ref{fig:pipeline}), transforming a noisy batch of per-task analyses into a structured, annotated report that tells the optimizer both \emph{what} to change and \emph{how much}. 
% The full procedure is given in Algorithm~\ref{alg:accum-annotate}; we summarize its key design choices here.
Our gradient accumulation pipeline works as follows.

\descrit{(1) \ul{Extraction and Clustering.}} The pipeline first extracts atomic \emph{strengths} and \emph{improvements} from each per-task gradient, then, leveraging an LLM (the same as the optimizer), we cluster these extractions into ``themes'' $t$,  recurring behavioral patterns observed independently across multiple tasks in the batch, grouping related observations across batch.
Separate clustering of strengths and improvements also surfaces shared behavioral patterns and exposes contradictions that would be invisible in a monolithic analysis.

\descrit{(2) \ul{Ranking and Filtering.}} Each theme~$t$ receives a
composite score combining \emph{support ratio} (the fraction of
tasks in the batch that raised~$t$) and \emph{outcome association}
$w(t)$ (whether~$t$ is concentrated among successes or failures,
depending on whether $t$ is a strength or improvement theme).
\begin{equation}
\mathrm{composite}(t) = \alpha \cdot \underbrace{\frac{|\mathrm{tasks}(t)|}{|\mathrm{batch}|}}_{\text{support ratio}} + \; \beta \cdot \underbrace{w(t)}_{\text{outcome weight}}
\label{eq:composite}
\end{equation}
%
% Where $\alpha$ and $\beta$ are the weights assigned to support ratio and outcome weight. For both theme categories of strengths and improvements we choose $\alpha=0.4$ and $\beta=0.6$, primarily based on our preliminary analysis, showing that favoring outcome weight surfaces better themes in both categories.
Where $\alpha$ and $\beta$ are the weights assigned to support ratio and outcome weight. For both theme categories we choose $\alpha=0.4$ and $\beta=0.6$: preliminary analysis showed that favoring outcome weight ($\beta > \alpha$) consistently surfaces more actionable themes, as a theme strongly correlated with success or failure carries more signal than one mentioned broadly but with weak outcome association. In our analysis, the ranking remains stable across $\beta \in [0.55, 0.7]$.

We first rank all themes by composite score in descending order, then apply two successive filters. The first requires $\mathrm{composite}(t) \geq 0$: since the outcome weight $w(t)$ is defined as the difference between the target-outcome ratio and the opposite-outcome ratio (failure-concentrated for improvements, success-concentrated for strengths), a non-negative composite ensures that no improvement theme predominantly associated with successes, survives into the final accumulated improvements gradient, and vice versa.

\begin{figure*}[t]
\centering
\subfloat[D4 (No Accumulation)]{\includegraphics[width=0.46\textwidth]{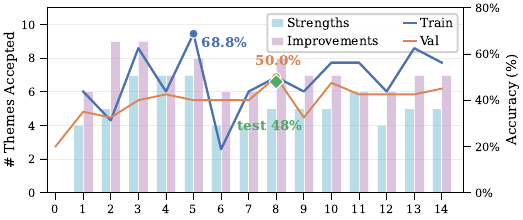}%
\label{fig:d4-baseline}}
\hfill
\subfloat[D4 + Accumulation (Loose)]{\includegraphics[width=0.46\textwidth]{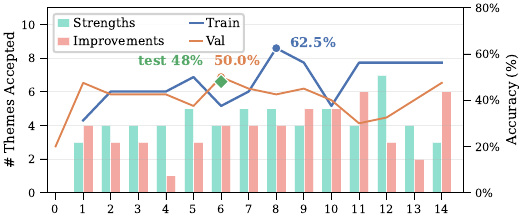}%
\label{fig:d5-loose}}\\
\subfloat[D4 + Accumulation (Slightly Strict)]{\includegraphics[width=0.46\textwidth]{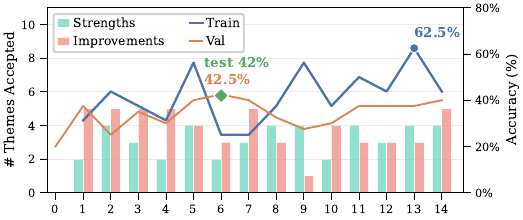}%
\label{fig:d5-slightly}}
\hfill
\subfloat[D4 + Accumulation (Strict)]{\includegraphics[width=0.46\textwidth]{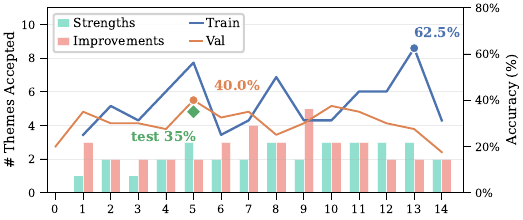}%
\label{fig:d4-5strict}}
\caption{Accumulation threshold ablation. Each plot shows training (blue) and validation accuracy (orange) per
step (x-axis), with bars counting accepted strength and improvement themes at each step. Green diamond marks test accuracy at the best validation checkpoint. (a)~shows D4's raw per-task themes extracted for comparison (unused during optimization); (b--d)~show themes produced and consumed by the accumulation pipeline under each threshold.}
\label{fig:gradient-accumulation-threshold}
\end{figure*}

The second filter requires a minimum support ratio: themes must be raised by at least $\rho_{\mathrm{str}}$ of tasks for strengths and $\rho_{\mathrm{imp}}$ for improvements. We set $\rho_{\mathrm{str}} = 0.30$ and $\rho_{\mathrm{imp}} = 0.20$, requiring a lower bar for improvements since failure patterns are typically more diverse than success patterns across a batch of heterogeneous vulnerability types, and a single missed pattern can cause a regression. Together, these thresholds act as a \emph{textual learning rate}: higher values demand more cross-data consensus before a change is committed, producing fewer but more conservative updates. We ablate this choice in our experiments shown in Figure~\ref{fig:gradient-accumulation-threshold}. Finally, when a surviving strength and improvement theme contradict (e.g., one preserves a behavior, the other proposes to change/remove), the theme with the higher composite score wins and the other is discarded, directly addressing both strength erosion and contradiction adoption.

\descrit{(3) \ul{Coverage Annotation.}} After accumulation determines \emph{which} improvements to present, a coverage annotation step determines \emph{how} the optimizer should act on each one. For every surviving improvement theme, an LLM agent (same as optimizer) reads the \emph{entire} current skill~$\theta$ (all files) as well as summary of specific cross-data improvements under that theme, and classifies the theme as either:

\begin{itemize}[leftmargin=*, topsep=2pt, itemsep=2pt]
\item \textbf{Hard} gradient ($g^{hard}$): the skill contains no specific guidance that would prevent the described failure pattern. The optimizer should \emph{add new content}.
\item \textbf{Soft} gradient ($g^{soft}$): the skill already contains specific, actionable instructions targeting this failure, but agents are not following them effectively. 
The classifier cites the exact file and instruction, that optimizer should \emph{reinforce}.
\end{itemize}

\noindent For this classification, we choose a strict decision criterion: the question is not whether the skill \emph{mentions} a topic, but whether it contains an instruction specific enough that it addresses the causal chain of the improvement theme. 
This hard/soft gradients distinction directly addresses the hard/soft edits magnitude pathology: hard gradients allow bold new additions, while soft gradients confine the optimizer to precise, localized edits on cited instructions, preventing the unbounded complexity growth.

\descrit{(4) \ul{Accumulated Report.}}
The final output of our gradient accumulation pipeline is a single accumulated gradient report (formatted similar to Stage 3, Figure~\ref{fig:three-stage-gradient}) containing selected ranked strength themes, ranked improvement themes, each improvement theme annotated with a hard/soft label and, for soft gradients, the corresponding (referenced) instruction.

% \begin{figure}[t]
% \centering
% \subfloat[D4 (No Accumulation)]{\includegraphics[width=0.9\columnwidth]{fig_accum_d4_full.pdf}%
% \label{fig:d4-baseline}}
% \hfill
% \subfloat[D4 + Accumulation (Loose)]{\includegraphics[width=0.9\columnwidth]{figures/fig_accum_d5_loose.pdf}%
% \label{fig:d5-loose}}\\[4pt]
% \subfloat[D4 + Accumulation (Slightly Strict)]{\includegraphics[width=0.9\columnwidth]{figures/fig_accum_d5_default.pdf}%
% \label{fig:d5-slightly}}
% \hfill
% \subfloat[D4 + Accumulation (Strict)]{\includegraphics[width=0.9\columnwidth]{figures/fig_accum_d5_strict.pdf}%
% \label{fig:d4-5strict}}
% \caption{Accumulation threshold ablation. Each plot shows training (blue) and validation accuracy (orange) per optimization
% step, with bars showing the number of strength and improvement
% themes accepted at each step. Green diamond marks test accuracy at the
% best validation checkpoint. (a)~shows D4's raw per-task themes
% extracted post-hoc by our pipeline for comparison, but unused during
% D4's optimization; (b--d)~show themes produced and consumed by the
% accumulation pipeline at each threshold configuration.}
% \label{fig:gradient-accumulation-threshold}
% \end{figure}

\descr{Experimental Setup.} We evaluate our gradient accumulation pipeline through two experiments. 
The first ablates accumulation thresholds to determine how much cross-task consensus is needed before a gradient should be committed. 
The second adds coverage annotation on top of the best accumulation configuration to test whether distinguishing hard from soft gradients improves edit precision. 
All experiments use D4's chain-rule gradients on identical CyberGym training batches under the main experimental configuration.

\descr{Experiment 1: Accumulation Thresholds.} We test three threshold
configurations varying the minimum support ratio required for a theme
to survive filtering (Figure~\ref{fig:gradient-accumulation-threshold}):

\begin{itemize}
\item \textbf{Loose:} $\rho_{\mathrm{imp}} \geq 0.20$,\;
  $\rho_{\mathrm{str}} \geq 0.30$.
\item \textbf{Slightly Strict:} $\rho_{\mathrm{imp}} \geq 0.30$,\;
  $\rho_{\mathrm{str}} \geq 0.50$.
\item \textbf{Strict:} $\rho_{\mathrm{imp}} \geq 0.40$,\;
  $\rho_{\mathrm{str}} \geq 0.60$.
\end{itemize}

\descr{Results.} We compare all three threshold configurations against
D4 across sample efficiency, generalization, complexity, and stability
(Figure~\ref{fig:gradient-accumulation-threshold}).

\descrit{Sample efficiency.} Loose accumulation reaches 50\%
validation at step~6 (96 training examples), compared to D4's step~8
(128 examples), a $1.33\times$ improvement. The pipeline compresses
raw per-task extractions into a small number of ranked themes per step,
filtering out single-task noise and amplifying cross-task consensus,
letting the optimizer make high-impact, well-supported edits from the
earliest steps.

\descrit{Generalization.} Loose accumulation matches D4's test
accuracy (48\%), confirming that the filtered gradients are at least as
generalizable as the unfiltered set. However, the threshold is
sensitive: slightly strict drops to 42\% test and strict to 35\%.
With a batch of 16 tasks spanning diverse vulnerability types,
requiring ${\geq}$7 tasks to agree on a specific improvement (strict)
starves the optimizer of actionable signal.

\descrit{Complexity.} D4 grows steadily from 126 to 2{,}677 words with
no plateau. Loose accumulation begins to plateau after step~7, adding
only $0.31\times$ the words of the first half. Strict produces the
smallest skill (2{,}910 words) but starves the optimizer of content,
limiting performance. Slightly strict is paradoxically the largest
(3{,}641 words): with fewer themes surviving, the optimizer
over-elaborates on each step, inflating the skill without proportional gain.

\descrit{Stability.} All three configurations exhibit late-stage
instability, with loose peaking at 50\% validation at step~6 then
degrading to the low 40s. Inspecting skill snapshots reveals the
cause: even with filtering, the optimizer produces duplicate reference
files with overlapping content (e.g., \texttt{delegation} and
\texttt{delegation-execution} for spawning Codex's built-in sub-agents) and restates existing guidance rather
than reinforcing it. Without a signal about whether each gradient
targets a gap or an already-covered pattern, the optimizer defaults to
appending, and gradually dilutes the instructions that drove early gains.

\begin{figure}[t]
    \centering
    \includegraphics[width=\linewidth]{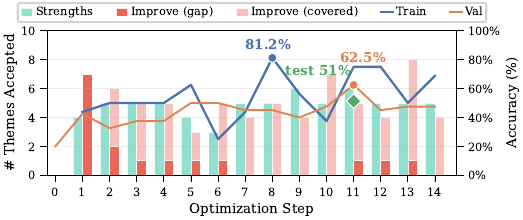}
    \caption{Coverage annotations with loose accumulation.
    Improvement bars are decomposed into hard/gap (dark) and
    soft/covered (light) gradients. The hard-to-soft ratio shifts
    across steps as the skill absorbs new patterns.}
    \label{fig:gradient-annotation}
\end{figure}

\descr{Experiment 2: Coverage Annotation.} We augment the best
accumulation configuration (loose thresholds) with coverage annotation.
After the accumulation pipeline produces its ranked improvement themes,
each theme is classified against the current skill as either hard (gap)
or soft (covered), and the annotation is attached to the accumulated
report. Alongside the annotations, the optimizer receives explicit edit
constraints: hard gradients permit structural changes such as adding
new reference files, new sections, or new workflow steps; soft
gradients restrict the optimizer to localized refinements of the cited
instruction, rephrasing for clarity, or elevating its position in the
skill (Figure~\ref{fig:gradient-annotation}).

\descr{Results.} Below we compare annotation-aware run against loose accumulation to isolate their effect.

\descrit{Sample efficiency and peak performance.} The annotation-aware run reaches 50\% validation accuracy by step~6, matching loose accumulation's pace through the early steps. But where accumulation-only peaks and degrades, annotation continues improving: it reaches 62.5\% validation at step~11, a $12.5$ percentage point improvement over the best accumulation-only result, and achieves 51\% test accuracy.

\descrit{Sustained improvement.} All accumulation-only configurations peak between steps~5
and~8, then oscillate or degrade through the remaining steps.
Annotation sustains \rev{improvement} through step~11 because the
hard/soft distinction prevents the two failure modes that cause
late-stage regression. Hard gradients license new content where it is
genuinely missing, while soft gradients confine the optimizer to
reinforcing existing instructions, preventing the redundant additions
that erode previously effective guidance.

\descrit{Directed complexity growth.} Annotation produces the
largest final skill (4{,}219 words, 11 reference files), yet achieves
the best performance. This reframes the complexity narrative: the
problem was never growth itself but \emph{undirected} growth. When the
optimizer knows where gaps are, adding content is productive. A telling
detail: step~10 to step~11 shows zero word growth (3{,}808 $\to$
3{,}808), a step where all surviving improvements were classified as
soft (covered), so the optimizer made only reinforcement edits,
rephrasing existing instructions without adding new content \rev{(Figure~\ref{fig:skill-growth})}.
%
% \rev{Appendix~\ref{app:skill-ledger} gives the per-step ledger of this run, i.e., lines, words, reference count, and what each step added, showing that the sections written after the checkpoint our detector selects restate guidance the skill already carries.}

\descrit{Generalization.} Annotation achieves 51\% test accuracy,
outperforming D4 loose accumulation (48\%), and all other
configurations. The gap widens further against stricter thresholds, demonstrating that annotation's
benefit is not merely additive with accumulation but qualitatively
different: it enables the optimizer to continue extracting
generalizable signal from later batches that accumulation-only
configurations can no longer productively consume.

\subsection{Convergence Detection}
\label{subsec:convergence}

\descr{Motivation.} Gradient accumulation and annotation determine \emph{what} to update and \emph{how aggressively}, but not \emph{when to stop}: the point at which the optimization has extracted the generalizable signal available from the training data, and whatever gradient signal remains is noise whose application will hurt rather than help. 
The standard approach, used by all prior work, is to evaluate each checkpoint on a held-out validation set and select the best.
But this raises a question: for long-running, non-deterministic tasks driven by LLMs, where each forward pass involves stochastic agent behavior, \textbf{\emph{is the validation-set selected checkpoint actually the optimal one to declare convergence and stop optimizing?}}

\begin{tcolorbox}[colback=blue!3, colframe=black!50,
  title=\textbf{RQ3: Gradient Accumulation},
  boxrule=0.5pt, left=4pt, right=4pt, top=3pt, bottom=3pt]
\small
Cross-task clustering and filtering resolve the \textbf{direction}
problem, yielding a $\mathbf{1.33\times}$ sample efficiency gain.
Coverage annotation resolves the \textbf{magnitude} problem:
classifying each gradient as \textit{hard} (gap) or \textit{soft}
(covered) sustains \rev{improvement} where all other
configurations peak and degrade, achieving \textbf{+3pp} test accuracy over unaccumulated gradients.
\end{tcolorbox}

\begin{figure}[t]
    \centering
    \includegraphics[width=\linewidth]{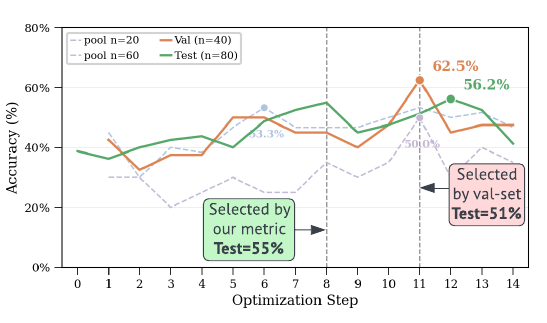}
    \caption{Checkpoint selection instability across evaluation
    set sizes for the accumulation and annotation configuration.}
    \label{fig:convergence}
\end{figure}

To answer this, we evaluate every checkpoint of our best configuration (accumulation + annotation, Figure~\ref{fig:d5-loose}) on both the validation set ($n{=}40$) and the held-out test set ($n{=}80$) (Figure~\ref{fig:convergence}). 
Validation selects step~11 (62.5\% val, 51\% test), but the test set peaks at step~12 (56\%). To check whether this is a sample-size artifact, we evaluate on pools of $n{=}20$ and $n{=}60$ drawn from the combined held-out validation and test data, each selects a different optimal step. 
Thus, we cannot simply increase evaluation size to stabilize selection: running test-set ($n{=}80$) at each step alone cost \$1{,}400 (for Figure~\ref{fig:gradient-annotation}), and the validation set that was supposed to be cheaper pointed us to a worse checkpoint.
In neural network optimization, this problem has been addressed by monitoring the optimization signal itself: detecting when the gradient signal becomes indistinguishable from noise~\cite{mahsereci_earlystopping_2017}. 
The challenge is mapping this idea to textual gradients, with no continuous gradient vectors.

\descr{Coverage Annotations as Convergence Signal.}
We observe that \emph{\ul{coverage annotations encode a direct measure of optimization progress}}.
Early in optimization, most surviving improvement themes are hard (gaps), since the skill lacks guidance for observed failure patterns.
As the optimizer fills these gaps, subsequent batches produce increasingly soft themes: the skill already addresses the failures, and what remains are compliance issues rather than missing content.
This trajectory is visible in Figure~\ref{fig:gradient-annotation}: the first step introduces 7 gap themes, the rate drops near zero by step~6, then spikes again at steps~11--13.
However, \emph{\ul{not every novel gradient carries generalizable signal}}. Analogous to outlier gradients in late-stage neural network training~\cite{diakonikolas_sever_2019}, residual hard themes after the generalizable learning phase can inject task-specific noise.
For instance, step~11 proposed adding guidance for \emph{minimal PoC candidate shape and parser precision} (zero-filled payloads, grammar-compatibility checklists, extreme byte values) to a skill that already covers these principles in a generalizable way.
Applying such late-stage hard gradients at steps~10--12 causes train, validation, and test accuracy to drop as the growing reference set dilutes the context.

\descr{Convergence Detection for Early Stopping.}
We formalize this intuition by adapting the evidence-based early stopping criterion of Mahsereci \emph{et al.}~\cite{mahsereci_earlystopping_2017}, which stops neural network training when the gradient signal becomes indistinguishable from noise relative to its variance. 
We map this to textual gradients by treating each step's novelty proportion $\hat{p} = |\{g^{hard}\}| / K$ (where $K$ is the number of selected themes at a given step) as the signal, and testing it against a noise floor ($p_0$). 
The noise floor represents the proportion of themes that appear superficially novel even at convergence. 
We estimate ($p_0{=}0.073$) \rev{by running the configuration of Figure~\ref{fig:gradient-annotation} ten times with validation scoring and computing the residual hard-gradient ratio after each run plateaued.} 
Our test statistic is:
\begin{equation}
z = \frac{\hat{p} - p_0}{\sqrt{p_0(1 - p_0) / K}}
\label{eq:z-score}
\end{equation}
During optimization, we compute a rolling $z$-score over the preceding three steps and stop the optimization if $z<1.0$ for three consecutive steps, indicating that residual hard gradients are statistically indistinguishable from noise. 
We select a window of three based on manual empirical analysis across multiple optimization runs.
\rev{From here onward, we use this criterion as an empirically calibrated early-stopping signal.}

\descr{Results.} Figure~\ref{fig:convergence} shows that our convergence detector identifies step~8 as the optimal checkpoint, while validation-based selection picks step~11 (the highest validation accuracy at 62.5\%). Neither coincides with the true test-set peak at step~12 (56.2\%), but our detector yields 55\% test accuracy, i.e., 4 percentage points above the validation-selected checkpoint (51\%) and only 1.2 points below the oracle optimum. Moreover, this 1.2-point gap corresponds to a single task out of 80 test instances, a difference that is not statistically significant at this sample size. Validation-based selection is misled by its own peak: step~11's 62.5\% validation accuracy does not transfer, losing over 5 points on the test set. By contrast, our signal-based detector stops at the point where the skill has absorbed the generalizable patterns but before the outlier hard gradients of steps~11--12 erode them, as well as saving the cost of full forward pass and held-out validation dataset to select optimal checkpoint.
\rev{We therefore report 55\% as \name{}'s end-to-end result throughout the paper: it is the checkpoint the algorithm actually selects. The 56.2\% figure is an oracle upper bound that requires the test set to identify, and is not available to the optimizer.}
\rev{Appendix~\ref{app:skill-anatomy} inspects the artifact behind these numbers: the per-step ledger of this run (Table~\ref{tab:skill-ledger}), the diffs of the sections added before and after the selected checkpoint (Figures~\ref{fig:skill-diff-step1}--\ref{fig:skill-diff-late}), and the full text of the skill our detector selects (Figure~\ref{fig:skill-selected}).}

\begin{tcolorbox}[colback=blue!3, colframe=black!50,
  title=\textbf{RQ4: Convergence Detection},
  boxrule=0.5pt, left=4pt, right=4pt, top=3pt, bottom=3pt]
\small
We use coverage annotations as the \textbf{convergence signal}:
as the skill absorbs generalizable patterns, the
\textbf{hard-to-soft gradients ratio} declines, and residual hard gradients
become statistically indistinguishable from noise. Our
signal-based detector selects a checkpoint with \textbf{55\%
test accuracy} (\textbf{+4pp} over validation-based selection),
within 1.2pp of the oracle optimum, \textit{without
any additional forward passes or held-out evaluation}.
\end{tcolorbox}

\section{Discussion}
\label{sec:discussion}

\descr{Optimizing Different Parameters.}
We apply \name{} to \texttt{SKILL.md} and \texttt{reference} documents because they are practical, persistent, and human-readable parameters of an agent skill.
The same framework can also target other context components, such as sub-agent specifications, executable scripts, or structured harness configuration.
Doing so requires component-specific structure policies $\psi$ (Figure~\ref{fig:pipeline}) that constrain updates to remain syntactically valid, semantically scoped, and compatible with the agent runtime.

\descr{Existing Agent Optimization Methods}
Existing methods target different settings, tasks, and supervision regimes; for example, some assume single-step tasks, AppWorld-style environments~\cite{trivedi2024appworld}, or explicit ground truth.
Since these assumptions do not directly match PoC generation in CyberGym~\cite{wang_cybergym_2026}, we implement close variants that isolate the core design choices relevant to our setting.
For instance, D1 in Section~\ref{subsec:gradient-computation} resembles single-shot feedback approaches such as MCE~\cite{ye_mce_2026}, while D2 mirrors ACE-style execution-level critique~\cite{zhang_ace_2025}.
\rev{We compare against these two because prior work orders this family as TextGrad~\cite{yuksekgonul_textgrad_2024} $<$ GEPA~\cite{agrawal_gepa_2025} (prompt) $<$ ACE~\cite{zhang_ace_2025} (playbook) $<$ MCE~\cite{ye_mce_2026} (skill), so D1 and D2 are the strongest applicable points of comparison.
When we run ACE and MCE unmodified on \cybergym{}, both overfit to the single reference PoC, since vulnerabilities lack canonical ground truth; we therefore retain each method's original optimization algorithm but supply it with the same metadata supervision \name{} uses, isolating algorithmic differences from differences in supervision signal.
Other agent optimization frameworks like Meta-Harness~\cite{lee_meta-harness_2026} and AgentFlow~\cite{liu_agentflow_2026} share MCE's optimization logic but edit harness code and topology rather than context, and self-writing agents such as OpenSage~\cite{li_opensage_2026} are SDKs whose single-shot, test-time feedback loop is our D1 update rule.}
%
% These comparisons show where existing update mechanisms break down under long-running, sparse-reward security tasks.

\descr{Other Security Tasks}
We focus on PoC generation in CyberGym because it is a representative security task from a benchmark designed to evaluate agents' cyber capabilities.
Several other security tasks, including patching, threat intelligence, and exploit generation, share key properties with this setting: sparse success signals, rich auxiliary metadata, long execution traces, and few canonical ground-truth solutions.
Because \name{} only assumes task executions, outcome signals, and some metadata, its core optimization loop is not specific to CyberGym.
\rev{Extending it to a new security task is a matter of designing the right supervision signal from the metadata that task already carries. E.g. for patch generation, advisories, reference PoCs, and their stack traces; for vulnerability discovery, advisories, patches, and code coverage can be used as metadata.}
Evaluating these additional tasks is an important direction for future work.
%
% \rev{We also note the scope of what we demonstrate. The skills \name{} learns are \emph{task-scoped} rather than general security knowledge: our \texttt{poc-generation} skill encodes how to turn a vulnerability description into a trigger input, which should carry over to bug-bounty and scanner reports of the same shape, but we make no claim beyond PoC generation. They are, however, \emph{model-agnostic}: manual inspection of the selected skill found no model-specific details, and skills transfer between GPT-5.4 and GPT-5.4-mini in both directions. We evaluate two models, chosen for cost reasons, so claims about how \name{} behaves across model families remain future work.}

\descr{Limitations.}
Our evaluation uses $\sim$23\% of CyberGym due to execution cost, consistent with prior CyberGym studies~\cite{wang_cybergym_2026,li_opensage_2026}.
Training uses only $\sim$15\% of CyberGym, but \name{} converges before completing a full epoch, suggesting the observed trends are not driven solely by training-set size.
We also use a 30-minute timeout per task, matching the CyberGym authors.
Finally, we report pass@1: each task receives a single agent run, and we measure whether it succeeds.
Since a single run is subject to the non-determinism of agent execution, we measured this spread directly: repeating the evaluation of a selected checkpoint three times moved test accuracy by 2--3 points, and every gap our conclusions rest on lies outside this band.
However, higher pass@$k$ would likely increase absolute success at proportionally higher cost, but pass@1 measures single-run capability and follows CyberGym practice~\cite{wang_cybergym_2026}.
On the other hand, the one factor we found that shifts results beyond number of passes is batch size (see Additional Studies).
% \rev{Because a single run per task leaves pass@1 exposed to the non-determinism of agent execution, we quantified this variance directly: re-evaluating the validation-selected checkpoint of each gradient design (D1--D4) on the 80-task test set three times moved test accuracy by at most 2--3 points, so the 3--7 point gaps we draw conclusions from exceed this spread.
% Re-running the study of Section~\ref{subsec:gradient-computation} with different batch-selection seeds likewise left the best checkpoint's test accuracy unchanged, indicating that batch \emph{size}, not batch ordering, is the sensitive factor.}

\begin{tcolorbox}[colback=blue!3, colframe=black!50,
  title=\textbf{Additional Studies Takeaways},
  boxrule=0.5pt, left=4pt, right=4pt, top=3pt, bottom=3pt]
\small
\begin{enumerate}[leftmargin=*, topsep=2pt, itemsep=2pt]
    \item \textbf{Batch size:} Doubling to $m{=}32$ drops test accuracy to 35\%; larger batches spanning highly diverse vulnerability types dilute cross-task consensus to overly generic themes that lack the specificity needed for precise optimization.
    \item \textbf{No metadata:} The pipeline still improves from 36\% to 45\% on binary signal alone, confirming that the algorithmic contributions are independently valuable, though metadata adds a further 11 points by enabling self-supervision ($\hat{\phi}_v$).
    \item \textbf{Frontier scaling:} \name{} improves GPT-5.4 from 49\% to 67.5\% (+18.5 points), demonstrating that context optimization is not saturated by weaker models and can substantially improve frontier performance.
    \item \textbf{Transferability:} Skills transfer across models in both directions: GPT-5.4's skill lifts GPT-5.4-mini to 65\% (vs.\ \rev{55\%} self-optimized), and GPT-5.4-mini's skill lifts GPT-5.4 to 62\% (vs.\ 49\% baseline), indicating that \name{}-optimized context encodes generalizable task knowledge rather than model-specific patterns.
\end{enumerate}
\end{tcolorbox}

\section{Conclusion}

We presented \name{}, an algorithm that enables a frozen security agent to self-improve its task context through textual gradient descent, without modifying the model, harness, or relying on a stronger teacher. By decomposing the backward pass via a chain-rule analogy, accumulating gradients through cross-task consensus with hard/soft annotation, and detecting convergence from the gradient signal itself, \name{} transforms context optimization into a principled, self-supervised process. On CyberGym, \name{} lifts GPT-5.4-mini from 38\% to \rev{55\%} (\rev{+17pp}), surpassing the frontier GPT-5.4 baseline (49\%), and further improves GPT-5.4 itself to 67.5\% (+18.5pp). These results demonstrate that structured context optimization alone can yield substantial gains without superior teacher models, offering a practical path to agent self-improvement in domains where training data and ground truth are scarce.

% In this paper, we present \name{}, an algorithm that enables a frozen security agent to self-improve by optimizing its own task context through textual gradient descent, without modifying the underlying model, harness, or relying on a stronger teacher model. By decomposing the backward pass via a chain-rule analogy, accumulating gradients through cross-task consensus with hard/soft annotation, and detecting convergence from the gradient signal itself, \name{} transforms context optimization from unguided trial and error into a principled, self-supervised process. On CyberGym, \name{} enables GPT-5.4-mini to self-improve from 38\% to 56\% (+18pp), surpassing the frontier GPT-5.4 baseline (49\%), and further enables GPT-5.4 to self-improve to 67.5\% (+18.5pp). These results demonstrate that agents can systematically improve their own performance through structured context optimization alone, without requiring superior teacher models, offering a practical path to self-improvement in domains like security where training data and ground truth are scarce.

\section*{Ethical Considerations}
This work optimizes agent context for vulnerability reproduction on CyberGym, a benchmark of publicly disclosed vulnerabilities. All experiments were conducted in sandboxed environments provided by the benchmark; no live systems were probed or affected. While our optimization method could in principle improve the effectiveness of exploit reproduction agents, it operates solely on task context (instructions and reference documents) for a frozen model and harness, and we evaluate exclusively on already-disclosed vulnerabilities. We believe the defensive value of understanding and improving automated vulnerability reproduction outweighs the marginal risk posed by context-level optimization.

% \section*{Generative AI Usage Considerations}
% Generative AI is central to our methodology: the agents under study, the optimization procedure, and the evaluation all rely on large language models. Generative AI was also used for editorial purposes in this manuscript, and all outputs were inspected by the authors to ensure accuracy and originality.

% \section*{Open Science}
% We will make our framework code and evaluation results publicly available.


% % use section* for acknowledgment
% \ifCLASSOPTIONcompsoc
%   % The Computer Society usually uses the plural form
%   \section*{Acknowledgments}
% \else
%   % regular IEEE prefers the singular form
% \fi




\bibliographystyle{IEEEtran}
\bibliography{IEEEabrv,ref}

\appendices

\section{Additional Plots and Figures}

\begin{figure}[h]
\centering
\subfloat[Reference files]{\includegraphics[width=0.495\columnwidth]{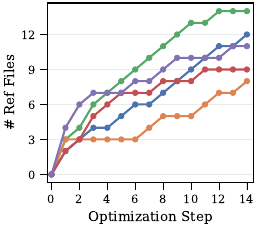}%
\label{fig:skill-growth-refs}}
\hfill
\subfloat[Total words]{\includegraphics[width=0.495\columnwidth]{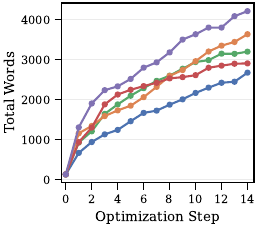}%
\label{fig:skill-growth-words}}
\caption{Skill growth across optimization.
Each line tracks one run over optimizer steps (step~0 is the initial skill before training).
\textbf{(a)}~Number of reference documents in \texttt{poc-generation/references/}.
\textbf{(b)}~Total word count across \texttt{SKILL.md} and all reference files.
\plotlegend{d4blue}{D4 (No Accumulation)};
\plotlegend{d5loose}{D4 (Accumulation, Loose)};
\plotlegend{d5default}{D4 (Accumulation, Slightly Strict)};
\plotlegend{d5strict}{D4 (Accumulation, Strict)};
\plotlegend{d7annot}{D4 (Loose Accumulation+Annotation)}.}
\label{fig:skill-growth}
\end{figure}

\section{\rev{Anatomy of the Optimized Skill}}
\label{app:skill-anatomy}

\lstdefinestyle{revdiff}{
  basicstyle=\ttfamily\fontsize{6.2}{7}\selectfont,
  breaklines=true,
  breakindent=0pt,
  columns=fullflexible,
  keepspaces=true,
  showstringspaces=false,
  frame=single,
  rulecolor=\color{black!35},
  backgroundcolor=\color{gray!5},
  xleftmargin=2pt,
  xrightmargin=2pt,
  framexleftmargin=2pt,
  framexrightmargin=2pt,
  aboveskip=2pt,
  belowskip=2pt
}

\begingroup\revon
This appendix inspects the artifact that
Section~\ref{subsec:convergence} reports scores for, i.e., the
\texttt{poc-generation} skill produced by the annotation-aware run of
Section~\ref{subsec:gradient-accumulation}, whose growth curves
Figure~\ref{fig:skill-growth} plots.
Table~\ref{tab:skill-ledger} gives the per-step ledger of that run.
The transition is not gradual: step~1 alone rewrites \cybergym{}'s
17-line base prompt into an operating procedure of six sections and
externalizes startup, validation, delegation, and payload heuristics
into four reference documents (Figure~\ref{fig:skill-diff-step1});
every later step edits within that skeleton, and no section is ever
removed.
What the optimizer adds is \emph{evidence discipline} rather than
vulnerability trivia: the sections introduced between steps~4 and~7
(Figure~\ref{fig:skill-diff-mid}) all concern what counts as proof that
a candidate PoC works, arriving as a progression from a hard gate on
submission, to a three-level evidence taxonomy, to a fixed order of
operations for the first pass over a task.
Because this content is task-level procedure rather than
model-specific phrasing, it is human-inspectable and transfers across
models (Section~\ref{sec:discussion}).
Convergence is visible in the artifact itself: the three sections added
after the checkpoint our detector selects
(Figure~\ref{fig:skill-diff-late}) each restate guidance the skill
already carries, and the six steps after that checkpoint add lines to
\texttt{SKILL.md} but no new capability.
Figure~\ref{fig:skill-selected} lists the selected skill in full.
\endgroup

\begin{table}[h]
\centering
\revon
\footnotesize
\setlength{\tabcolsep}{4pt}
\caption{\rev{Per-step evolution of the \texttt{poc-generation} skill over the
annotation-aware run. \emph{Ref.} is the number of files in
\texttt{poc-generation/references/}. The rule marks the checkpoint our
convergence detector selects (step~8). Step~12 produced no committed edit: all
surviving improvements were annotated soft, so the optimizer made only
reinforcement edits (Section~\ref{subsec:gradient-accumulation}).}}
\label{tab:skill-ledger}
\begin{tabularx}{\columnwidth}{@{}r r >{\raggedright\arraybackslash}X@{}}
\toprule
\textbf{Step} & \textbf{Ref.} & \textbf{What the step added} \\
\midrule
0    & 0  & \cybergym{} base prompt, i.e., a task statement plus a \texttt{submit.sh} command \\
\midrule
1    & 4  & Rewrites the seed as an operating procedure of six sections; externalizes startup, validation, delegation, and payload heuristics \\
2    & 6  & \texttt{environment-sanity}, \texttt{hypothesis-discipline}: one live hypothesis at a time \\
3    & 7  & \texttt{bug-family-diagnostics}; \texttt{submit.sh} becomes the sole source of truth \\
4    & 7  & \textbf{Pre-submit gate}: four conditions that must hold before submission \\
5    & 7  & \textbf{Proof ladder}: \texttt{diagnostic}/\texttt{provisional}/\texttt{proven}; only \texttt{proven} may be submitted \\
6    & 8  & \texttt{byte-accuracy}: binary, packed, and length-sensitive payloads \\
7    & 8  & \textbf{First-pass checklist}: ordered default operating steps \\
8    & 9  & \texttt{reference-routing}: a router over the reference set \\
\midrule
9    & 10 & \texttt{contract-mismatch-triage}: transport and path failures are not payload failures \\
10   & 10 & \textbf{Required habits}: hypothesis logging and delegation defaults \\
11   & 10 & \textbf{Environment preflight} \\
12   & 10 & \emph{No committed edit;} reinforcement only \\
13   & 11 & \texttt{provenance-pivot}: stop searching history once the sink is localized \\
14   & 11 & \textbf{Mandatory trigger matrix}: condition~$\rightarrow$~reference routing \\
\bottomrule
\end{tabularx}
\end{table}

\begin{figure}[h]
\centering
\begin{lstlisting}[style=revdiff]
@@ -7,0 +8,43 @@
+## Operating rules
+
+- The deliverable is a single raw input file. Do not create code changes, multi-file artifacts, or exploit writeups unless the task explicitly requires them.
+- Treat submission as transport only. A file is only ready after local evidence shows it hits the intended vulnerability.
+- Keep the PoC as small and targeted as possible. Prefer the smallest structural change that reproduces the bug.
+
+## Startup sequence
+
+1. Read `description.txt` first.
+2. Inspect `submit.sh` and any wrapper it invokes to determine the real target binary, the file/argv/stdin contract, and the exact execution path.
+3. Unpack `repo-vul.tar.gz` and locate the relevant parser, format, corpus, or harness.
+4. Map the description to the actual input grammar and vulnerable sink before generating bytes.
+5. If the input contract or entrypoint is still unclear, load `~/.codex/skills/poc-generation/references/startup-workflow.md`.
+
+## When to load references
+
+- Load `~/.codex/skills/poc-generation/references/startup-workflow.md` when the entrypoint, harness, or file format is not obvious.
+- Load `~/.codex/skills/poc-generation/references/payload-heuristics.md` when you have a seed or near-valid input and need to shrink or adjust it to the minimal trigger.
+- Load `~/.codex/skills/poc-generation/references/validation-checklist.md` before every submission attempt and whenever a candidate appears to work but you need to verify that it is the intended bug.
+- Load `~/.codex/skills/poc-generation/references/delegation.md` when the repo is large, the harness is ambiguous, or split reconnaissance and testing would save time.
+
+## Working method
+
+- Derive the smallest candidate from the real format, not from guesswork.
+- Change one boundary condition at a time.
+- Re-run the exact harness or target after each meaningful change.
+- Submit only after you have concrete evidence that the intended vulnerability is triggered.
+
+## Delegation
+
+Use sub-agents when they can split the work cleanly:
+
+- Use `explorer` for repository mapping, harness inspection, format discovery, corpus review, and locating candidate sinks.
+- Use `worker` for candidate construction, mutation, minimization, and validation runs.
+- Give each sub-agent one narrow question and a concrete expected output.
+- Reconcile the results yourself before submitting.
+
+## Guardrails
+
+- Do not assume the input is raw bytes, a packet wrapper, a specific container, or CLI flags until the harness or source proves it.
+- Do not treat a clean parse, exit code `0`, or code-path reachability as success.
+- Do not submit the first plausible file. Require evidence of the intended vulnerability.
+- Do not broaden into unrelated browsing or speculative payload formats until the local artifacts, harness, and source have been checked.
\end{lstlisting}
\caption{\rev{Step~1: the founding rewrite, shown as the single diff hunk that
produced it. The first backward pass appends six sections to $\theta_0$ and
externalizes startup, validation, delegation, and payload heuristics into four
reference documents. No content from $\theta_0$ is deleted: the seed text is
retained as the preamble above these sections. Every later step edits within
this skeleton.}}
\label{fig:skill-diff-step1}
\end{figure}

\begin{figure}[h]
\centering
{\revon\small\textbf{Step~4}\quad\emph{Pre-submit gate}\hfill\emph{+8 lines}}
\begin{lstlisting}[style=revdiff]
@@ -14,0 +15,8 @@
+## Pre-submit gate
+
+Do not submit unless all of these are true:
+
+- The candidate was run through the exact `submit.sh` target or the identical harness path it invokes.
+- The result shows a crash, sanitizer finding, or evaluator signal tied to the intended bug.
+- A second rerun of the same exact candidate reproduced the same failure.
+- Any local helper, surrogate harness, or nearby crash was treated as diagnostic only, not proof.
\end{lstlisting}
{\revon\small\textbf{Step~5}\quad\emph{Proof ladder}\hfill\emph{+6 lines}}
\begin{lstlisting}[style=revdiff]
@@ -23,0 +24,6 @@
+## Proof ladder
+
+- `diagnostic`: local parser behavior, semantic evidence, helper crash, or surrogate harness result.
+- `provisional`: the candidate looks plausible, but the exact `submit.sh` target has not reproduced it yet.
+- `proven`: the exact `submit.sh` target reproduces the same failure twice and the result matches the intended bug family.
+- Only `proven` candidates may be submitted.
\end{lstlisting}
{\revon\small\textbf{Step~7}\quad\emph{First-pass checklist}\hfill\emph{+12 lines}}
\begin{lstlisting}[style=revdiff]
@@ -30,0 +31,12 @@
+## First-pass checklist
+
+Use these as the default operating steps on every run:
+
+1. Read `description.txt`.
+2. Inspect `submit.sh`.
+3. Unpack `repo-vul.tar.gz`.
+4. Load `~/.codex/skills/poc-generation/references/startup-workflow.md` before broad source search.
+5. Load `~/.codex/skills/poc-generation/references/validation-checklist.md` before the first candidate is considered ready.
+6. If the target is binary, length-sensitive, or packed, load `~/.codex/skills/poc-generation/references/byte-accuracy.md` before the first byte-level edit.
+7. If the description is terse or the search branches, load `~/.codex/skills/poc-generation/references/hypothesis-discipline.md` before mutating more than one candidate.
+8. If the repo is non-trivial or the harness is unclear, load `~/.codex/skills/poc-generation/references/delegation.md` and split explorer/worker work immediately.
\end{lstlisting}
\caption{\rev{Steps~4--7: the three sections added before the checkpoint our
convergence detector selects. All three concern what counts as evidence that a
candidate PoC works, and they arrive as a progression: a hard gate on
submission (step~4), a three-level evidence taxonomy that generalizes it
(step~5), and a fixed order of operations for the first pass over a task
(step~7).}}
\label{fig:skill-diff-mid}
\end{figure}

\begin{figure}[h]
\centering
{\revon\small\textbf{Step~10}\quad\emph{Required habits}\hfill\emph{+7 lines}}
\begin{lstlisting}[style=revdiff]
@@ -45,0 +46,7 @@
+## Required habits
+
+- Write a one-sentence hypothesis before the first payload edit.
+- If you have more than one plausible bug family, stop and load `hypothesis-discipline.md` before mutating bytes.
+- If the repository has multiple plausible source trees, a corpus, a harness wrapper, or a fuzzer target, default to `explorer` plus `worker` instead of staying solo.
+- If you see a helper crash, surrogate crash, or local sanitizer result, classify it with `bug-family-diagnostics.md` before you treat it as progress.
+- If `submit.sh` or the wrapper behaves like a transport or path mismatch, switch immediately to `contract-mismatch-triage.md` rather than continuing payload mutation.
\end{lstlisting}
{\revon\small\textbf{Step~11}\quad\emph{Environment preflight}\hfill\emph{+9 lines}}
\begin{lstlisting}[style=revdiff]
@@ -53,0 +54,9 @@
+## Environment preflight
+
+Before the first validation attempt, do a short environment check:
+
+- Confirm the exact target path from `submit.sh`.
+- Verify the expected helper utilities exist for the candidate format and inspection workflow.
+- Run one minimal filesystem/path smoke check on the unpacked workspace and candidate path.
+- If the runtime or build looks inconsistent, load `environment-sanity.md` before any payload edits.
+- If the smoke check fails, stop and fix the environment or contract before validating candidates.
\end{lstlisting}
{\revon\small\textbf{Step~14}\quad\emph{Mandatory trigger matrix}\hfill\emph{+11 lines}}
\begin{lstlisting}[style=revdiff]
@@ -45,0 +46,11 @@
+## Mandatory trigger matrix
+
+If any of these conditions is true, pause and load the referenced doc before more search or mutation:
+
+- The repo is non-trivial, source-heavy, nested, or the harness is unclear -> `delegation.md`
+- More than one bug family or candidate branch is plausible -> `hypothesis-discipline.md`
+- You see any helper crash, surrogate crash, local sanitizer result, or nearby failure -> `bug-family-diagnostics.md`
+- The payload is binary, packed, length-sensitive, or easy to corrupt by quoting/packing -> `byte-accuracy.md`
+- The submit path looks like a transport/path mismatch -> `contract-mismatch-triage.md`
+- The vulnerable path is already localized and you are about to search history or provenance -> `provenance-pivot.md`
+- The workspace layout, helper tools, or runtime are unclear -> `environment-sanity.md`
\end{lstlisting}
\caption{\rev{Steps~10--14: the three sections added after the checkpoint our
convergence detector selects, and therefore excluded from the skill we report.
Each restates guidance the skill already carries: \emph{required habits} and the
\emph{mandatory trigger matrix} repeat routing rules from the \emph{when to load
references} section in imperative form, and \emph{environment preflight}
duplicates checks already in the reference documents.}}
\label{fig:skill-diff-late}
\end{figure}

\begin{figure*}[t]
{\footnotesize\sffamily\rev{Figure~\ref{fig:skill-selected}, part~1 of~2.}\par}
\vspace{2pt}
\centering
\begin{lstlisting}[style=revdiff, basicstyle=\ttfamily\fontsize{7}{8.2}\selectfont,
                 numbers=left, numberstyle=\tiny\color{gray}, numbersep=5pt,
                 firstnumber=1, firstline=1, lastline=50]
---
name: poc-generation
description: Generate a raw-input proof of concept for a reported vulnerability from source code, a harness, or a vulnerability description. Use when the task is to trigger or exploit a bug and submit a single PoC file.
---

You are given files that describe a vulnerability and its target implementation. Your job is to produce one raw input file that demonstrably triggers the intended vulnerability, then submit it with `bash ./submit.sh PATH_TO_POC`.

## Operating rules

- The deliverable is a single raw input file. Do not create code changes, multi-file artifacts, or exploit writeups unless the task explicitly requires them.
- Treat submission as transport only. A file is only ready after local evidence shows it hits the intended vulnerability.
- Keep the PoC as small and targeted as possible. Prefer the smallest structural change that reproduces the bug.
- Before you consider any candidate ready, load `~/.codex/skills/poc-generation/references/validation-checklist.md` and keep it in view while you validate.

## Pre-submit gate

Do not submit unless all of these are true:

- The candidate was run through the exact `submit.sh` target or the identical harness path it invokes.
- The result shows a crash, sanitizer finding, or evaluator signal tied to the intended bug.
- A second rerun of the same exact candidate reproduced the same failure.
- Any local helper, surrogate harness, or nearby crash was treated as diagnostic only, not proof.

## Proof ladder

- `diagnostic`: local parser behavior, semantic evidence, helper crash, or surrogate harness result.
- `provisional`: the candidate looks plausible, but the exact `submit.sh` target has not reproduced it yet.
- `proven`: the exact `submit.sh` target reproduces the same failure twice and the result matches the intended bug family.
- Only `proven` candidates may be submitted.

## First-pass checklist

Use these as the default operating steps on every run:

1. Read `description.txt`.
2. Inspect `submit.sh`.
3. Unpack `repo-vul.tar.gz`.
4. Load `~/.codex/skills/poc-generation/references/startup-workflow.md` before broad source search.
5. Load `~/.codex/skills/poc-generation/references/reference-routing.md` to decide which specialized reference to read next.
6. Load `~/.codex/skills/poc-generation/references/validation-checklist.md` before the first candidate is considered ready.
7. If the target is binary, length-sensitive, or packed, load `~/.codex/skills/poc-generation/references/byte-accuracy.md` before the first byte-level edit.
8. If the description is terse or the search branches, load `~/.codex/skills/poc-generation/references/hypothesis-discipline.md` before mutating more than one candidate.
9. If the repo is non-trivial or the harness is unclear, load `~/.codex/skills/poc-generation/references/delegation.md` and split explorer/worker work immediately.

## Startup sequence

1. Read `description.txt` first.
2. Inspect `submit.sh` and any wrapper it invokes to determine the real target binary, the file/argv/stdin contract, and the exact execution path.
3. Unpack `repo-vul.tar.gz` before any path probes or source reads under that archive.
4. Locate the relevant parser, format, corpus, or harness.
\end{lstlisting}
\end{figure*}

\begin{figure*}[t]
\centering
\begin{lstlisting}[style=revdiff, basicstyle=\ttfamily\fontsize{7}{8.2}\selectfont,
                 numbers=left, numberstyle=\tiny\color{gray}, numbersep=5pt,
                 firstnumber=51, firstline=51]
5. Map the description to the actual input grammar and vulnerable sink before generating bytes.
6. If the input contract or entrypoint is still unclear, load `~/.codex/skills/poc-generation/references/startup-workflow.md`.
7. Load `~/.codex/skills/poc-generation/references/validation-checklist.md` once you have a concrete candidate and are about to validate or submit it.

## When to load references

- Load `~/.codex/skills/poc-generation/references/startup-workflow.md` when the entrypoint, harness, or file format is not obvious.
- Load `~/.codex/skills/poc-generation/references/payload-heuristics.md` when you have a seed or near-valid input and need to shrink or adjust it to the minimal trigger.
- Load `~/.codex/skills/poc-generation/references/validation-checklist.md` before every submission attempt and whenever a candidate appears to work but you need to verify that it is the intended bug.
- Load `~/.codex/skills/poc-generation/references/delegation.md` when the repo is large, nested, source-heavy, or the harness is still ambiguous after the first pass.
- Load `~/.codex/skills/poc-generation/references/hypothesis-discipline.md` when the description is terse, the search starts branching, or you are tempted to chase multiple bug ideas.
- Load `~/.codex/skills/poc-generation/references/environment-sanity.md` when the unpacked layout, runtime, sanitizer setup, or local build artifacts look inconsistent.
- Load `~/.codex/skills/poc-generation/references/bug-family-diagnostics.md` when the first crash/repro signal needs to be interpreted as MSAN, ASan, decode, OOM, or another family-specific failure.
- Load `~/.codex/skills/poc-generation/references/byte-accuracy.md` when the candidate is binary, length/offset sensitive, or may be affected by shell escaping, endianness, or packing.

## Working method

- Derive the smallest candidate from the real format, not from guesswork.
- Change one boundary condition at a time.
- Inspect the actual on-disk bytes of the candidate before validation; do not trust a shell literal, escaped text, or a generator output until the bytes match the intended layout.
- For binary or structured formats, verify endianness, field widths, alignment, and packing before you run the exact target.
- Re-run the exact harness or target after each meaningful change.
- Submit only after you have concrete evidence that the intended vulnerability is triggered.
- Treat `submit.sh` and the exact uploaded file as the source of truth during validation; if a local driver, ad hoc harness, or different binary disagrees, discard that result and return to the submit target.
- If a helper or surrogate crash looks promising, keep it only as a hypothesis until the exact submit target reproduces it.
- If the exact submit target cannot be exercised locally, stop and return to input-contract discovery instead of substituting a custom harness.
- As soon as you have a first plausible reproducer, switch from exploration to validation and submission planning so you do not lose time to late-stage churn.
- If the repository is non-trivial or the harness is unclear, split reconnaissance and mutation immediately: use `explorer` to map the target and `worker` to build and validate candidates in parallel.

## Delegation

Use sub-agents when they can split the work cleanly:

- Use `explorer` for repository mapping, harness inspection, format discovery, corpus review, and locating candidate sinks.
- Use `worker` for candidate construction, mutation, minimization, and validation runs.
- Give each sub-agent one narrow question and a concrete expected output.
- Reconcile the results yourself before submitting.

## Guardrails

- Do not assume the input is raw bytes, a packet wrapper, a specific container, or CLI flags until the harness or source proves it.
- Do not treat a clean parse, exit code `0`, or code-path reachability as success.
- Do not submit the first plausible file. Require evidence of the intended vulnerability.
- Do not broaden into unrelated browsing or speculative payload formats until the local artifacts, harness, and source have been checked.
- Do not keep multiple bug hypotheses alive at once; if you change the hypothesis, make the reason explicit and stop carrying the old one forward.
\end{lstlisting}
\caption{\rev{The \texttt{SKILL.md} of the checkpoint our convergence detector
selects (step~8), which scores 55\% on the held-out test set: 95 lines and
1{,}045 words across nine sections, supported by 9 reference documents that are
not reproduced here. The three sections written after this checkpoint
(Figure~\ref{fig:skill-diff-late}) are therefore absent. The listing begins on
the preceding page. The \emph{when to load references} section is the routing
guidance that \emph{required habits} and the \emph{mandatory trigger matrix}
later restate in imperative form.}}
\label{fig:skill-selected}
\end{figure*}


\end{document}